\documentclass[twocolumn,aps,prb,longbibliography]{revtex4-1}
\usepackage[dvips]{graphicx}
\usepackage{epsfig} 
\usepackage{color}
\usepackage[normalem]{ulem}
\usepackage[labelformat=simple]{subfig}
\usepackage{amsmath,amsfonts,amssymb,bm}
\usepackage{caption}
\usepackage{bm}
\usepackage{cancel}
 \newcommand{\SpinUp}{{\color{red} \big\uparrow}}
  \newcommand{\SpinDown}{{\color{blue} \big\downarrow}}
\UseRawInputEncoding
\setcitestyle{numbers,square}

\begin{document}
\title{Exact solution of the minimalist Stark many body localization problem  in terms of spin  pair hopping.}
\author{Alexander L. Burin}
\affiliation{Tulane University, New
Orleans, LA 70118, USA}

\date{\today}
\begin{abstract}
Simultaneous conservation of charge (spin projection to the $z$ axis) and dipole moment can partially suppress ergodicity (thermalization) by means of shattering the phase space. This shattering results in many body localization  of some states even in the absence of disordering, while other states remain delocalized. Here we show for the minimalist one-dimensional spin $1/2$ model how to distinguish localized and delocalized states, based on  any representative product state with given projections of spins to the $z$ axis,  separating states into four groups characterized by distinguishable behaviors. These include  two groups of delocalized states  with translationally invariant Krylov subspaces  with integrable   (group {\bf I}) or ergodic  (group {\bf II})  dynamics,    and  the other two groups with confined spin transport having  either all mobile spins (group {\bf III}) or some immobile (frozen)  spins (group {\bf IV}). The states of the first two groups are delocalized, while the states of the last two groups are mostly localized. The theory is used to interpret recent experiments  \cite{Guo2020MBLTiltedModel,Monroe2021observationStarkMBL} and suggest their extension necessary to observe both localized and delocalized behaviors in a dipole moment conserving regime.
\end{abstract}



\maketitle

\section{Introduction}
\label{sec:intr}

Ergodicity breakdown represents the failure of quantum statistical mechanics \cite{Deutsch91ETH,Srednicki94ETH,Lukin2016DisorderFreeLoc} and gives advantages for quantum information processing in many-body systems \cite{PhysRevX.4.011052,Bahri2015,PhysRevB.89.144201}. It takes place in the form of many body localization (MBL) due to strong disordering \cite{Gornyi05,Basko06,OganesyanHuse07,Imbrie16} similarly to the single particle Anderson localization \cite{Anderson58,FleishmanAnderson80} or can be possibly realized without disordering \cite{KaganMaksimov84,MIKHEEV83He3He4,KaganMaksimov85,ab90Kontor,
Logan,LeitnerArnoldDiffusion97,Shepelyansky98,Carleo2012,Roeck14,
Muller15DisordFreeLoc,AbaninDisFree15,
Lukin2016DisorderFreeLoc,
Michailidis18SlowDynRegSys,Brenes18MBLReg,Serbyn2021,
Kovrizhin17LocalDisFree,Kovrizhin17bLocalDisFree,Brenes18MBLReg, Lukin2016DisorderFreeLoc, vanNieuwenburg2019StarMBL1st,Pollmann2019StarkMBL2nd, Huse2DinFieldPhysRevX}. Localization without disorder is an exciting fundamental task since it conflicts with the common sense expectations. 
It is also   an sttractive experimental challenge  for researchers in a quantum information area \cite{Guo2020MBLTiltedModel,Monroe2021observationStarkMBL} because of the localization robastness. Indeed, the localization in a regular system  is insensitive to a specific disorder realization. 

Disorder free localization was considered at high temperature due to thermal disorder in positions of interacting particles \cite{KaganMaksimov84}. It emerges  in a thermodynamic limit of infinite system if some particles are static \cite{Kovrizhin17LocalDisFree,Kovrizhin17bLocalDisFree}; yet it is unstable with respect to arbitrary small deviation from a static behavior \cite{Lukin2016DisorderFreeLoc,Kovrizhin2018Gaugeprb}. However, the localization  can be robust in systems conserving not only number of particles but dipole moment  or higher moments 
\cite{Pollmann20Fragment,Moudgalya20QHall,Moudgalya2019thermalization,Rakovszky2020, Khemani20Fragmentation,HuseFreezingDipCons, Bardarson21FragmentedSpaceSampleSt,Pal08-2021Spin1Shattering,
DeTomasiprl20SlowTrasp,GShepelyanskyStarkClassic2009,Feldmeier21SlowDyn, GornyiPolyakov21ShatteringNumerics,Zakrzevskii20MBLHarmTrap}. Dipole moment  is approximately conserved  in  fractional quantum Hall effect   \cite{Bergholtz05QHallXYmodel,SeidelQHall05,Wang12QHallDipCons,Moudgalya20QHall} and systems subjected to a large potential energy gradient compared to the bandwidth (Stark MBL, Refs.  \cite{vanNieuwenburg2019StarMBL1st,Pollmann2019StarkMBL2nd, Huse2DinFieldPhysRevX,Bloch21StarkMBL}). 
On the one hand some eigenstates in those systems are completely {\it frozen}, i. e.  they are characterized by  fixed spin projections to the $z-$axis  \cite{Pollmann20Fragment}, while on the other hand there exist other fully delocalized ergodic eigenstates \cite{Turner2018,Moudgalya20QHall,Moudgalya2019thermalization,
Bardarson21FragmentedSpaceSampleSt}. 
Eigenstate  behavior (localized or delocalized) in systems with local, dipole moment conserving hopping \cite{Pollmann20Fragment,Moudgalya2019thermalization,Rakovszky2020,
Moudgalya20QHall,Bardarson21FragmentedSpaceSampleSt}  depends on any representative product state  determining the unique Krylov subspace of all product states coupled to that state by the system Hamiltonian. It was demonstrated that different families of Krylov subspaces exist with different dynamical properties, including localizing and non-localizing dynamics.  {\it  These findings  \cite{Pollmann20Fragment,Moudgalya2019thermalization,Rakovszky2020,
Moudgalya20QHall,Bardarson21FragmentedSpaceSampleSt}  motivate us to seek for the determination  of eigenstate properties using their representative product states that is the primary target of the present work.}  

The product states are  usually chosen in the experiments as the initial states
\cite{Guo2020MBLTiltedModel,Monroe2021observationStarkMBL,Bloch21StarkMBL} giving us the opportunity to realize any regime of interest by choosing the proper initial state. However, the full localization was reported in the dipole moment conserving regime of  a large field gradient  \cite{Guo2020MBLTiltedModel,Monroe2021observationStarkMBL} independent of the initial state.  Our second target is {\it to interpret these observations and suggest ways to attain the full diversity of  behaviors}. 

In this work we establish the unique connection between the basis product states and the localization of eigenstates  in the minimalist, dipole moment conserving periodic model (the MM model) on a chain defined as (cf. Refs. \cite{NandkishoreFractCirc2019,Bloch21StarkMBL,
Moudgalya20QHall,Moudgalya2019thermalization})  
\begin{eqnarray}
\widehat{H}_{MM}=\Delta\sum_{k=1}^{N}(S_{k}^{+}S_{k+1,p}^{-}S_{k+2,p}^{-}S_{k+3,p}^{+}+H.C.) 
\nonumber\\
+2\Delta\sum_{k=1}^{N}S_{k}^{z}(S_{k+1,p}^{z}-S_{k+2,p}^{z}),
\label{eq:MinimHamCons}
\end{eqnarray}
where $S_{k,p}^{a}=S_{k}^{a}$ for $k\leq N$ and  $S_{k,p}^{a}=S_{k-N}^{a}$ for $k> N$ ($a=+$,  $-$ or $z$). The dipole moment $\widehat{P}=\sum_{k=1}^{N}S_{k}^{z}(k-(N+1)/2)$ is conserved in this model  with the accuracy to an integer number of $N'$s (modulo $N$) \cite{Bardarson21FragmentedSpaceSampleSt}. 

The present model is defined as the outcome of the parent $XY$ model with the nearest and next neighbor interactions subjected to the large field gradient  that is approximately relevant to the systems investigated in Refs. \cite{Guo2020MBLTiltedModel,Monroe2021observationStarkMBL}. In Sec. \ref{sec:ModelDeriv} we show that the periodicity in space in Eq. (\ref{eq:MinimHamCons}) can be attained applying transverse ($XY$ model) and longitudinal (Stark field) interactions consecutively and periodically in time. Eq. (\ref{eq:MinimHamCons})  is derived using the generalized Schrieffer-Wolff transformation  \cite{Schreiber15ScienceExp,DynkinEtcSTRICHARTZ1987320,Achilles2012,ab15MBLXY,
ab19Quartets}, in the lowest non-vanishing order in inverse field gradient $F$.   The violation of spatial periodicity of the parent model results in  additional longitudinal fields emerging in the lower order in $F$. These fields function as quenched disorder  causing the  localization of all states at sufficiently large field gradient observed experimentally \cite{Guo2020MBLTiltedModel,Monroe2021observationStarkMBL} as  discussed in Sec. \ref{sec:Exp}. 

In Sec. \ref{sec:LocDeloc}  we investigate the Krylov subspaces of product spin states of the model Eq. (\ref{eq:MinimHamCons}) and introduce the four groups of states distinguished by their dynamic behavior as confirmed by the analysis of group averaged imbalances. In Sec. \ref{sec:Exp} the experimental data of Refs. \cite{Guo2020MBLTiltedModel,Monroe2021observationStarkMBL} are discussed in light of our findings, The work ends by the extended conclusion and discussion section \ref{sec:Concl}, where the results of the present work are briefly resumed, the   comparison of them with the earlier work is outlined and the generalization to other models is discussed. The long derivations are placed  to the Appendix. 

\section{Psarent model and derivation of minimalist model} 
\label{sec:ModelDeriv}

The analytical results of the present work are mostly related to the minimalist periodic dipole moment conserving model, given by Eq. (\ref{eq:MinimHamCons}). This model represents the first non-vanishing expansion term  of the generalized Schrieffer-Wolff transformation \cite{Schreiber15ScienceExp,DynkinEtcSTRICHARTZ1987320,Achilles2012,ab15MBLXY,
ab19Quartets} of the parent $XY$ model subjected to the uniformly growing field in the large field gradient limit. Below we derive the Hamiltonian Eq. (\ref{eq:MinimHamCons}) for both open and periodic boundary conditions (OBC and PBC). The former case describes the  experiments \cite{Guo2020MBLTiltedModel,Monroe2021observationStarkMBL}, while the latter case represents their desirable generalization that realizes a spatial periodicity insensitive to boundaries, which are dramatically important for the delocalization in a large field gradient limit (see Sec. \ref{sec:Exp}).   The derivation below is not related to dynamic  properties of the model Eq. (\ref{eq:MinimHamCons}) considered in Sec. \ref{sec:LocDeloc} so those readers, who are interested only in the analysis of this model, can skip it. 

The minimalist model (MM) in Eq. (\ref{eq:MinimHamCons}) differs from the model of Refs. \cite{Moudgalya2019thermalization,Bardarson21FragmentedSpaceSampleSt} referred here as the minimalist hopping only (MH) model by the presence of the longitudinal term containing $S^{z}$ operators. As it is shown in the present section this term necessarily emerges as the outcome of the Schrieffer-Wolff transformation of the parent $XY$ model subjected to the strong field gradient.  Longitudinal and transverse interactions are of the same order of magnitude. 

The other parent models including  that oif the  fractional quantum Hall effect in  the thin-torus limit   \cite{Bergholtz05QHallXYmodel,SeidelQHall05,Wang12QHallDipCons,Moudgalya20QHall}  or 
the anisotropic Heisenberg model with nearest neighbor interactions subjected to a large  field gradient  \cite{Bardarson21FragmentedSpaceSampleSt,Pollman2021StarkMBLOpen,Taylor20ExpShatt}
 also lead to a  significant longitudinal interactions within the effective Hamiltonian in addition to the MH model Hamiltonian of Refs.  \cite{Moudgalya2019thermalization,Bardarson21FragmentedSpaceSampleSt}.  In those models longitudinal interactions exceed the transverse ones  in contrast with Eq. (\ref{eq:MinimHamCons}), where longitudinal and transverse interactions are comparable. 
  Longitudinal interaction does not modify Krylov subspaces of relevant product states. However, it  affects spin dynamics in those states, as discussed  in Sec. \ref{sec:ConclLoc}, enhancing  the localization.  
  

\subsection{Effective Hamiltonian of dipole moment conserving system with open boundary conditions.}
\label{sec:ModelOBC}


We begin with the derivation of the effective Hamiltonian for the general $XY$ model with open boundary conditions subjected to a field uniformly increasing   by a certain gradient $F$ between adjacent sites. This model can be characterized by the Hamiltonian $\widehat{H}$ expressed as the sum of the field ($\widehat{H}_{F}$) and $XY$ model ($\widehat{H}_{XY}$ )Hamiltonians  
\begin{eqnarray}
\widehat{H}=\widehat{H}_{F}+\widehat{H}_{XY}, ~ \widehat{H}_{F}=-F\sum_{k=1}^{N}\left(k-\frac{N+1}{2}\right)S_{k}^{z}, 
\nonumber\\
 \widehat{H}_{XY}=\frac{1}{2}\sum_{i< j}^{N}J_{ij}(S_{i}^{+}S_{j}^{-}+S_{j}^{-}S_{i}^{+}). 
\label{eq:Hn}
\end{eqnarray}
This model, referred as the parent  OBC model,  characterizes both transmon qubits within the superconducting quantum processor, investigated in Ref. \cite{Guo2020MBLTiltedModel}, where the interaction is limited to nearest and next neighbors, and  pseudospin states of  interacting $^{171}$Yb$^{+}$ ions,  investigated in Ref.  \cite{Monroe2021observationStarkMBL}, where the interaction depends on the distance as $J_{ij}=J/|i-j|^{1.3}$. The minimalist model  Eq. (\ref{eq:MinimHamCons})  is the outcome of the  Schrieffer-Wolff transformation of the parent model  Eq. (\ref{eq:Hn}) in the large field gradient limit $F > J_{ij}$ with the only  nearest and next neighbor interactions ($J_{1}$ and $J_{2}$, respectively) different from zero. 
The minimalist $XY$ model relevant for the many body localization problem cannot be restricted to only nearest neighbor interactions because this model is equivalent to non-interacting fermions \cite{JordanWigner28}. 

 
In a large field gradient limit $F\gg J$ the effective Hamiltonian projected to the subspace of states with identical dipole moments can be derived using the Schrieffer-Wolff transformation similarly to Refs. \cite{ab15MBLXY,Moudgalya2019thermalization}. The zeroth order Hamiltonian is the longitudinal field Hamiltonian $\widehat{H}_{F}$ and the perturbation is given by the $XY$ model Hamiltonian $\widehat{H}_{XY}$ in Eq. (\ref{eq:Hn}). Since any term in the perturbation does not conserve the dipole moment and thus modifies the zeroth order energy by a large field gradient $F$ the Schrieffer-Wolff transformation is well justified at large field gradients $F \gg J_{ij}$.

Applying the Schrieffer-Wolff transformation  we modify the original Hamiltonian  as 
\begin{eqnarray}
\widehat{H}\rightarrow e^{\widehat{S}}\widehat{H}e^{-\widehat{S}},
\label{eq:SW1}
\end{eqnarray}
where $\widehat{S}$ is an anti-Hermitian matrix chosen to eliminate the perturbation Hamiltonian $\widehat{H}_{XY}$ in the first order in $\widehat{S}$ that requires $[\widehat{S}, \widehat{H}_{F}]=-\widehat{H}_{XY}$. This is sufficient to derive the effective Hamiltonian in a desirable third order, where the minimalist model  Eq. (\ref{eq:MinimHamCons}) emerges.    

The matrix  $\widehat{S}$ is defined in terms of razing and lowering spin operators $S_{k}^{\pm}=S_{k}^{x}\pm iS_{k}^{y}$ as 
\begin{eqnarray}
\widehat{S}=\frac{1}{2}\sum_{i< j}^{N}\frac{J_{ij}}{F(j-i)}(S_{i}^{+}S_{j}^{-}-S_{j}^{-}S_{i}^{+}).
\label{eq:SSchrWf}
\end{eqnarray}
The expansion of the effective Hamiltonian Eq. (\ref{eq:SW1}) in $\widehat{S}$ up to the third order term yields  \cite{SchriefferWolff66} 
\begin{eqnarray}
\widehat{H} \approx \widehat{H}_{F}+\left\{\frac{1}{2}\left[\widehat{S}, \widehat{V}\right]+\frac{1}{3}\left[\widehat{S}, \left[\widehat{S}, \widehat{V}\right]\right]\right\}_{P},
\label{eq:SW2}
\end{eqnarray}
where the subscript $P$ in the definition of the effective Hamiltonian $\widehat{H}_{eff}$ means that the only terms conserving dipole moment $P$ (modulo $N$) are  left. 
The third order term is left together with the second order one since the spin-spin interaction  in the $XY$ model appears only in the third order in $\widehat{V}$ \cite{ab15MBLXY,ab19Quartets}. 

The second order term can be expressed as \cite{ab15MBLXY} 
\begin{eqnarray}
\widehat{H}_{2}=\frac{1}{2}\left[\widehat{S}, \widehat{V}\right]=
\-\frac{1}{4F}\sum_{i}S_{i}^{z}\sum_{j\neq i}\frac{J_{i j}^2}{(i-j)}.
\label{eq:sw2nddiag}
\end{eqnarray}
Eq. (\ref{eq:sw2nddiag}) introduces a site dependent longitudinal  field acting on each spin. For the interaction determined by the interspin distance only, i. e. $J_{ij}=J(|i-j|)$ this term disappears in the macroscopic limit of an infinite number of spins because the sum over $j$ is anti-symmetric. 
This is the case for the  cold ions considered in Ref. \cite{Monroe2021observationStarkMBL} in the limit $N\rightarrow \infty$, while the interaction of transmon qubits in Ref. \cite{Guo2020MBLTiltedModel} does not satisfy the rule $J_{ij}=J(|i-j|)$ since it vanishes for certain pairs of next neighbor spins. 

The interaction Eq. (\ref{eq:sw2nddiag})  is significant for both experiments \cite{Guo2020MBLTiltedModel,Monroe2021observationStarkMBL}  under consideration and it is responsible for the localization observed experimentally at different initial states  for large field gradients as shown in  Sec. V  below. However, this second order term disappears in periodic settings as shown in  Sec. \ref{sec:ModelPBC}. Therefore it is ignored in Eq. (\ref{eq:MinimHamCons})  where the periodic model is considered. 


The transverse hopping interaction violating the dipole moment conservation also emerges in the second order in $JF^{-1}$  in the form 
\begin{eqnarray}
\widehat{H}_{2offd}=\frac{1}{16F}\sum_{j\neq l, k}\frac{(j+l-2k)J_{jk}J_{kl}S_{k}^{z}S_{j}^{+}S_{l}^{-}}{(j-k)(l-k)}. 
\label{eq:sw2ndoffdiag}
\end{eqnarray}
 An additional Schrieffer-Wolff transformation is needed to eliminate it. This transformation  will generate dipole moment conserving   interactions of order of $J^4F^{-3}$,  which is smaller in the large field gradient limit  compared to the interactions  described by the second term in the expansion Eq. (\ref{eq:SW2}). 
This term contains both diagonal binary spin interaction \cite{ab15MBLXY} and transitions in spin quartets \cite{ab19Quartets}. The diagonal interaction takes the form 
\begin{eqnarray}
\widehat{H}_{3d}
=\frac{1}{2F^2}\sum_{j<k}S_{j}^{z}S_{k}^{z}\sum_{l}\frac{J_{jk}J_{kl}J_{lj}}{(k-l)(j-l))}.  
\label{eq:sw3rddiag}
\end{eqnarray}
If the hopping interaction is limited to nearest neighbors, i. e.  the only interaction $J_{i,i+1}=J_{1}$ differs from zero, then all interactions in Eq. (\ref{eq:sw3rddiag}) are equal zero \cite{ab15MBLXY} due to a single particle nature of the $XY$ model with nearest neighbor interactions \cite{JordanWigner28}. In the minimalist $XY$ model with nearest and next neighbor hopping interactions ($J_{i,i+1}=J_{1}$, $J_{i,i+2}=J_{2}$)  Eq. (\ref{eq:sw3rddiag}) generates nearest and next neighbor interactions in the form $\sum_{i<j}U_{ij}S_{i}^{z}S_{j}^{z}$ with interaction constants $U_{ij}$ defined as  
\begin{eqnarray}
U_{ij}=\frac{J_{1}^2J_{2}}{4F^2}\left(\delta_{i,j-1}(2-\delta_{i1}-\delta_{jN})-\delta_{i, j-2}\right). 
\label{eq:3rdDiagMinim}
\end{eqnarray}
Thus the nearest and next neighbor longitudinal interactions $\pm \Delta$, respectively, with $\Delta=J_{1}^2J_{2}/(2F^2)$ are generated for all sites except for those at the edges (cf. Eq. (\ref{eq:MinimHamCons})) where the nearest neighbor interaction is smaller by the factor of $2$. 

The off-diagonal four-spin hopping interaction conserving dipole moment can be evaluated similarly to Ref. \cite{ab19Quartets}  as 
\begin{eqnarray}
\widehat{H}_{3offd}=\sum_{i<j<k<l}V_{ijkl}\delta_{i+l,j+k}S_{i}^{+}S_{j}^{-}S_{k}^{-}S_{l}^{+}
\label{eq:sw3rdoffdiag}
\end{eqnarray}
where $\delta_{ab}$ is the Kronecker symbol  and the four spin interaction $V_{ijkl}$ can be expressed as
\begin{eqnarray}
J_{ijkl}=-\frac{1}{4F^2}\left(\frac{J_{ij}J_{ik}J_{il}}{(i-j)(i-k)}+\frac{J_{il}J_{jl}J_{kl}}{(l-j)(l-k)}
\right. 
\nonumber\\
\left.+ \frac{J_{ij}J_{jk}J_{jl}}{(j-i)(j-k)}+\frac{J_{ik}J_{jk}J_{kl}}{(k-i)(k-l)}\right).
\label{eq:4spinint}
\end{eqnarray}
If the only nearest neighbor interaction $J_{i,i+1}=J_{1}$ differs from zero the four spin hopping amplitude disappears 
\cite{ab19Quartets} that is the consequence of a single particle nature of $XY$ model with nearest neighbor interactions \cite{JordanWigner28}. For the simplest many-body $XY$ model  with nonzero nearest and next neighbor interactions ($J_{i,i+1}=J_{1}$, $J_{i,i+2}=J_{2}$)  Eq. (\ref{eq:4spinint}) leads to a non-zero hopping for only consecutive  spin quartets $i, i+1, i+2, i+3$.  For those quarters the only two last  terms in Eq. (\ref{eq:4spinint}) contribute to the four spin hopping amplitude $\Delta=J_{i,i+1,i+2,i+3}=J_{1}^2J_{2}/(4F^2)$ (cf. Eq. (\ref{eq:MinimHamCons})).

\subsection{Case of periodic boundary conditions.}
\label{sec:ModelPBC}

Here we derive the effective Hamiltonian Eq. (\ref{eq:MinimHamCons})  for the periodic model with a finite number of spins that is the main subject of study for the present work. {\it For the first time} we suggest the realization of the fully periodic Stark MBL problem for a finite number of spins  insensitive to the boundary conditions that can be realized experimentally. 

 To make hopping periodic  within the experimental settings  of Refs.  \cite{Guo2020MBLTiltedModel,Monroe2021observationStarkMBL} one should  connect transmon qubits  used in Ref. \cite{Guo2020MBLTiltedModel} periodically or place cold ions investigated in Ref. \cite{Monroe2021observationStarkMBL}  equidistantly in a circle. This should make spin hopping amplitudes in Eq. (\ref{eq:Hn}) periodic, i. e. $J_{ij}=J_{i+N,j}=J_{i,j+N}$.  A uniformly increasing field can be made periodic on a circle using  time periodic Hamiltonian  
defined as (cf. Eq. (\ref{eq:Hn}))
\begin{eqnarray}
\widehat{H}(t)=
\begin{cases}
\widehat{H}_{F} & \text{for } 2k\tau_{F}<t<(2k+1)\tau_{F},\\    
  \widehat{H}_{XY}  & \text{for }(2k+1)\tau_{F}<t<2(k+2)\tau_{F}, 
\end{cases}
\nonumber\\
\tau_{F}=\frac{2\pi}{NF}, ~ k=0, \pm 1, \pm 2,...
\label{eq:PerH}
\end{eqnarray}
The time period $2\tau_{F}$ is chosen to make the   interaction associated with the field gradient $F$  periodic on a circle. This periodicity can be understood considering  the system evolution matrix describing the Hamiltonian Eq. (\ref{eq:PerH}) action  during a period as
\begin{eqnarray}
\widehat{U}=\widehat{U}_{ph}\widehat{U}_{XY}, ~
 \widehat{U}_{ph}=e^{-i\sum_{k=1}^{N}\frac{2\pi k}{N}S_{k}^{z}}=e^{-i\widehat{S}_{0}},  
\nonumber\\
 \widehat{U}_{XY}=e^{-\sum_{k\neq j}\frac{2\pi i J_{kj}}{NF}S_{k}^{+}S_{j}^{-}}=e^{-i\widehat{S}_{1}}. 
\label{eq:pulsetot}
\end{eqnarray}
 In a large field gradient limit this evolution matrix allows conservation of the dipole moment $P=\sum_{k=1}^{N}\left(k-\frac{N+1}{2}\right)S_{k}^{z}$ with the accuracy to the integer number of spin numbers $N$ since the change of the dipole moment by $N$ modifies the evolution matrix $\widehat{U}_{ph}$ by $e^{2\pi i}=1$. This is exactly   the change of the dipole moment by $N$  in  the four spin hopping involving  edges (e. g. $S_{N-2}^{-}S_{N-1}^{+}S_{N}^{+}S_{1}^{-}$) that makes this hopping periodic on a circle  \cite{Bardarson21FragmentedSpaceSampleSt}. The periodicity on a circle of the effective  Hamiltonian conserving dipole moment  emerges naturally during  its construction as derived below. 

To be more specific we define periodicity on a circle for a certain operator  as following. For any multispin interactions within this operator  that can be generally expressed as $J_{i_{1},i_{2},...i_{n}}^{\alpha_{1},\alpha_{2},...\alpha_{n}}S_{i_{1}}^{\alpha_{1}}S_{i_{2}}^{\alpha_{2}}...S_{i_{n}}^{\alpha_{n}}$ ($\alpha_{i}=\pm, z$) the conditions of translational invariance and periodicity on a circle must be satisfied. These conditions can be written as  (respectively)
\begin{eqnarray}
J_{i_{1}+a,i_{2}+a,...i_{n}+a}^{\alpha_{1},\alpha_{2},...\alpha_{n}}= J_{i_{1},i_{2},...i_{n}}^{\alpha_{1},\alpha_{2},...\alpha_{n}},  ~ a=1,2,3..., 
\nonumber\\  
J_{i_{1},i_{2},...i_{k}+N...i_{n}}^{\alpha_{1},\alpha_{2},...\alpha_{n}}= J_{i_{1},i_{2},...i_{k},...i_{n}}^{\alpha_{1},\alpha_{2},...\alpha_{n}}, ~1\leq k \leq N.
\label{eq:TransInvPer}
\end{eqnarray}
The effective Hamiltonian derived by means of the generalized Schrieffer-Wolff transformation Eq. (\ref{eq:PulseSW})  possesses the property Eq. (\ref{eq:TransInvPer}) as shown below in Sec. \ref{sec:GenSchrWfTd}.

The period $2\tau_{F}$ of the Hamiltonian $\widehat{H}(t)$ Eq. (\ref{eq:PerH}) is the minimum time needed to make  a finite system periodic. For this specific choice in the limit of a large field gradient $F$ the approximate conservation  of the dipole moment (modulo $N$) takes place as needed to support a periodic spin hopping, see Ref. \cite{Bardarson21FragmentedSpaceSampleSt}. The period chosen as any integer multiple of $2\tau_{F}$ ($2k\tau_{F}$ with an integer, non-zero  $k$) also  ensures the  spatial periodicity. Yet, in that case the dipole moment is conserved  with the accuracy to an addition of integer numbers of $N/k$.  This can lead to  appearance  of more terms in the effective Hamiltonian conserving the dipole moment. For instance if $k=2$ and $N$ is even the hopping terms like $S_{k}^{+}S_{k+N/2}^{-}$ modifying the system dipole moment by $N/2$ will be also allowed. Additional hopping can reduce the localization contrary to the original goal to attain it in the maximum extent.  Therefore we use the minimum possible period $2\tau_{F}$. 

\subsubsection{Transformation of the evolution matrix}
\label{sec:GenSchrWfTd}

To eliminate  off-resonant processes and create the effective Hamiltonian conserving quasi-dipole moment $P$ (modulo  $N$)  
 or $e^{2\pi i \sum_{k=1}^{N} (k-(N+1)/2)S_{k}^{z}/N}$ \cite{Bardarson21FragmentedSpaceSampleSt}, one can apply a generalized Schrieffer-Wolff transformation \cite{SchriefferWolff66}  to the evolution matrix Eq. (\ref{eq:pulsetot}) for a period   in the form
\begin{eqnarray}
\tilde{U}_{*}=e^{\widehat{S}}\widehat{U}e^{-\widehat{S}} 
\label{eq:PulseSW}
\end{eqnarray}
with the antihermitian matrix $\widehat{S}$  (wave functions should be modified simultaneously as $\psi \rightarrow e^{\widehat{S}}\psi$).  

The goal of the transformation is to bring the evolution matrix to the form 
\begin{eqnarray}
\widehat{U_{*}}=\widehat{U}_{ph} e^{-i\widehat{H}_{eff}\tau_{F}},  
\label{eq:PulseTarget}
\end{eqnarray}
with the effective Hamiltonian $\widehat H_{eff}$ conserving the dipole moment (modulo $N$). This goal can be approximately attained eliminating the leading (in $JF^{-1}$) interactions modifying dipole moment in the system Hamiltonian using the generalized  Schrieffer-Wolf transformation. Then the  Baker-Campbell-Hausdorff-Dynkin formula \cite{DynkinEtcSTRICHARTZ1987320,Achilles2012} should be applied to express the action of modified evolution matrices in terms of   a single effective Hamiltonian $\widehat H_{eff}$. This new iteration for the effective Hamiltonian  will have  interactions violating the dipole moment conservation smaller by a factor of $J/F$ compared to the original Hamiltonian   
similarly to the standard  Schrieffer-Wolff transformation \cite{SchriefferWolff66} as illustrated below in Sec. \ref{sec:MinSpPer}. This procedure should be repeated an infinite number of times to get rid of the off-resonant interaction in all orders in $JF^{-1}$.  It should converge at large field gradients $F$ \cite{BakCampHausdConv20} as confirmed by our numerical calculations reported below in Sec. IV. 

We expect that the effective  Hamiltonian in Eq. (\ref{eq:PulseTarget}) is periodic on a circle as defined by  Eq. (\ref{eq:TransInvPer}). To show that one can represent the transformed evolution matrix introducing the operator $ \widehat{S}_{mod}$ as 
\begin{eqnarray}
\tilde{U}_{*}=e^{\widehat{S}}\widehat{U}_{ph}\widehat{U}_{XY}e^{-\widehat{S}}
=
\widehat{U}_{ph}e^{\widehat{S}_{mod}}\widehat{U}_{XY}e^{-\widehat{S}}, 
\nonumber\\
 \widehat{S}_{mod}=\widehat{U}_{ph}^{-1}\widehat{S}\widehat{U}_{ph}. 
\label{eq:TransEvMatr}
\end{eqnarray}
The Hamiltonian $\widehat{H}_{XY}$ and the associated evolution matrix $\widehat{U}_{XY}$ are periodic on a circle. Assume that this is true for the operator $\widehat{S}$ and this operator also conserves the projection of spin to the $z$ axis as the original Hamiltonian. The latter is obviously true for the standard Schrieffer-Wolff transformation. 

Then the same is true for the operator $\widehat{S}_{mod}$. Indeed, if the operator $\widehat{S}$ depends on operators $\{S_{n}^{\alpha}\}$ ($\alpha=\pm, z$, $n=1,2,...N$) then the operator $\widehat{S}_{mod}$ is the identical function of those operators, modified as $S_{n}^{\pm} \rightarrow S_{n}^{\pm} e^{\pm 2\pi in/N}$, $S_{n}^{z} \rightarrow S_{n}^{z}$. For the modified operator the property of translational invariance (the first line in Eq. (\ref{eq:TransInvPer})) is valid because of the identical number of $S^{+}$ and $S^{-}$ operators in each term as required by the spin projection conservation. Therefore, the additional factors $ e^{\pm 2\pi ia/N}$ (where $a$ is the shift of indices in Eq. (\ref{eq:TransInvPer})) compensate each other. The periodicity with the period $N$ is satisfied since the addition of the number of spins $N$ to the spin position $n$ in the exponents $e^{\pm 2\pi in/N}$ accompanying  operators $S_{n}^{\pm}$ modifies them as  $e^{\pm 2\pi i(n+N)/N}$ thus multiplying them by  $e^{\pm 2\pi i}=1$. 
The construction of the operator  $\widehat{S}$ is demonstrated below in Sec. \ref{sec:MinSpPer}. 

\subsubsection{Effective Hamiltonian in the leading (third) order of $1/F$ expansion. }
\label{sec:MinSpPer}

To find the effective Hamiltonian $\widehat H_{eff}$  in the first non-vanishing order in $J_{ij}/F$ one can use the transformation eliminating the  interaction  $\widehat{S}_{1}$ violating the dipole moment conservation and generating dipole moment conserving  terms. To find such transformation we introduce a new operator $\widehat{S}_{mod}$ as (cf. Eq. (\ref{eq:TransEvMatr}))
\begin{eqnarray}
e^{\widehat{S}}e^{-i\widehat{S}_{0}}=e^{-i\widehat{S}_{0}}e^{\widehat{S}_{mod}}.  
\label{eq:PulseSW-2}
\end{eqnarray}
Then the modified evolution matrix can be expressed in the form 
\begin{eqnarray}
\tilde{U}=e^{-i\widehat{S}_{0}}e^{\widehat{S}_{mod}}e^{-i\widehat{S}_{1}}e^{-\widehat{S}}. 
\label{eq:PulseSW-3}
\end{eqnarray}
The exponent $e^{-i\widehat{S}_{1}}$ gets canceled in the first non-vanishing order in $\widehat{S}_{1}$ if 
\begin{eqnarray}
\widehat{S}_{mod}-i\widehat{S}_{1}-\widehat{S}=0.   
\label{eq:PulseSW-4}
\end{eqnarray}

Seeking the operator $\widehat{S}$ in the general form of the sum of  binary products of spin raising and lowering operators 
\begin{eqnarray}
\widehat{S}=\sum_{k, l}A_{kl}S_{k}^{+}S_{l}^{-}    
\label{eq:SOper}
\end{eqnarray}
one can express the modified exponent as
\begin{eqnarray}
\widehat{S}_{mod}=\sum_{k, l}A_{kl}S_{k}^{+}S_{l}^{-} e^{2\pi i\frac{k-l}{N}}.  
\label{eq:SMod}
\end{eqnarray}
The amplitudes $A_{kl}$ satisfying Eq. (\ref{eq:PulseSW-4}) are  defined as
\begin{eqnarray}
A_{kl}=-\frac{1}{2NF}\frac{2\pi i J_{kl}}{1-e^{2\pi i\frac{k-l}{N}}}. 
\label{eq:Amps}
\end{eqnarray}

This definition of the amplitudes  $A_{kl}$ satisfies the requirement $A_{lk}=-A_{kl}^{*}$ so  the transformation of the evolution matrix in Eq. (\ref{eq:PulseSW}) is unitary as it is supposed to be. In the limit of large $N$ the operator $\widehat{S}$ becomes fully identical to the operator corresponding to the Schrieffer-Wolff transformation in Eq. (\ref{eq:SSchrWf}) for open boundary conditions.

Using this definition of the transformation one can represent the evolution matrix Eq. (\ref{eq:PulseSW})  in the form of Eq. (\ref{eq:PulseTarget}) as 
\begin{eqnarray}
\tilde{U}\approx e^{-i\widehat{S}_{0}}e^{-iH_{eff}\tau_{F}},
\label{eq:PMod}
\end{eqnarray}
with the effective Hamiltonian containing  only resonant terms conserving dipole moment (modulo $N$). The effective Hamiltonian can be obtained using the  Baker-Campbell-Hausdorff-Dynkin formula \cite{Achilles2012} up to the third order in  $JF^{-1}$  in the form 
\begin{widetext}
\begin{eqnarray}
\widehat{H}_{eff}=\left[\widehat{H}_{*}\right]_{P}, ~ \widehat{H}_{*}=
\frac{1}{i\tau_{F}}\left(-\widehat{S}_{mod}+i\widehat{S}_{1}+\widehat{S}\right)+
\frac{i}{2i\tau_{F}}\left(i[\widehat{S}_{mod}, \widehat{S}_{1}]+ [\widehat{S}_{mod}, \widehat{S}]-i[\widehat{S}_{1}, \widehat{S}]\right)
\nonumber\\
+\frac{i}{12\tau_{F}}\left(-i[\widehat{S}_{mod}, [\widehat{S}_{mod}, \widehat{S}_{1}]]+[\widehat{S}_{1}, [\widehat{S}_{mod}, \widehat{S}_{1}]]-3i[\widehat{S}, [\widehat{S}_{mod}, \widehat{S}_{1}]]\right)
\nonumber\\
+\frac{i}{12\tau_{F}}\left(-i[\widehat{S}_{mod}-
i\widehat{S}_{1}, [\widehat{S}_{mod}-i\widehat{S}_{1}, \widehat{S}]]-[\widehat{S}, [\widehat{S}_{mod}-i\widehat{S}_{1}, \widehat{S}]]\right), 
\label{eq:BCHExp}
\end{eqnarray}
\end{widetext}
where the subscript $P$ in the definition of the effective Hamiltonian $\widehat{H}_{eff}$ means that the only terms conserving dipole moment $P$ (modulo $N$) are  left. 
Using Eq. (\ref{eq:PulseSW-4}) one can simplify the Hamiltonian expressing it in terms of $\widehat{S}_{mod}$ and $\widehat{S}$ operators as (particularly, the first term and  the last line vanish because $\widehat{S}_{mod}-i\widehat{S}_{1}=- \widehat{S}$)
\begin{eqnarray}
\widehat{H}_{*}=
\frac{i}{2\tau_{F}}[\widehat{S}_{mod}, \widehat{S}]
\nonumber\\
+\frac{i}{6\tau_{F}}\left([\widehat{S}_{mod}, [\widehat{S}_{mod}, \widehat{S}]]+[\widehat{S}, [\widehat{S}_{mod}, \widehat{S}]]\right)
\label{eq:BCHExp-1}
\end{eqnarray}

Consider the second order term (the first term in Eq. (\ref{eq:BCHExp-1}), which is of the second order in $J/F$ compared to the leading Stark term, that is proportional to $F$). This term being  projected to the subspaces of the Hilbert space with identical dipole moments is represented by  spin dependent longitudinal fields  that can be expressed as (cf. Eq. (\ref{eq:sw2nddiag}))
 \begin{eqnarray}
\widehat{H}_{2}=-\frac{\pi}{4NF}\sum_{j, k}J_{kj}^2\cot(\pi (k-j)/N)S_{k}^{z}.
\label{eq:H2per2}
\end{eqnarray}
If the interaction $J_{ij}$ is periodic on a circle Eq. (\ref{eq:TransInvPer}), this term vanishes because the sum over $j$ is antisymmetric. This is the significant advantage of the periodic (PBC) model  compared to its OBC counterpart, where this second order term dramatically suppresses the delocalization creating effective disorder (see Sec. V). 


The transverse term  emerging  in the same order in $J_{ij}/F$ does not conserve the dipole moment and 
can be neglected similarly to the analogous term for the  OBC problem  Eq. (\ref{eq:sw2ndoffdiag}), when considering the contributions  to the effective Hamiltonian up to the third order in $JF^{-1}$. However,  it contributes to the fourth order interactions discussed    in Sec. \ref{sec:ModRel}.

The third order (in $JF^{-1}$) dipole-moment conserving contributions to the effective Hamiltonian comes from longitudinal  and transverse  terms  
\begin{eqnarray}
\widehat{H}_{eff}=\widehat{H}_{3,l}+\widehat{H}_{3,tr}. 
\label{eq:Heff3rd}
\end{eqnarray}
The longitudinal  term representing  the induced longitudinal spin-spin interactions can be evaluated  similarly to Ref. \cite{ab15MBLXY} and Eq. (\ref{eq:sw3rddiag})  as
\begin{widetext}
\begin{eqnarray}
\widehat{H}_{3,l}=\sum_{j<k}U_{jk}S_{j}^{z}S_{k}^{z}, ~ U_{jk}=
\frac{\pi^2}{6N^2F^2}\frac{J_{jk}J_{lj}J_{kl}\left(2\cos\left(\frac{\pi(j-k)}{N}\right)+\cos\left(\frac{\pi(2l-j-k)}{N}\right)\right)}{\sin\left(\frac{\pi(l-j)}{N}\right)\sin\left(\frac{\pi(l-k)}{N}\right)}.
\label{eq:EffHD}
\end{eqnarray}
\end{widetext}

The transverse  part of the effective Hamiltonian Eq. (\ref{eq:Heff3rd})  can be evaluated similarly to the OBC case of Eqs. (\ref{eq:sw3rdoffdiag}),  (\ref{eq:4spinint})   as 
\begin{widetext}
\begin{eqnarray}
\widehat{H}_{3,offd}=\sum_{j,k,l,m}V_{jklm}S_{j}^{+}S_{k}^{-}S_{l}^{+}S_{m}^{-}\Delta_{N}(j+l-k-m), ~
 \Delta_{N}(a)=\sum_{p=-\infty}^{\infty}\delta_{a, pN}, 
\label{eq:EffHOffD-resper}
\end{eqnarray}
\end{widetext}
where $\delta_{ab}$ is the Kronecker symbol. The interaction  $V_{ijkl}$ is defined as
\begin{widetext}
\begin{eqnarray}
 V_{ijkl}=-\frac{\pi^2}{12N^2F^2}
 \left(\frac{J_{ij}J_{ik}J_{il}\left[2\cos\left(\frac{\pi(i-j)}{N}\right)\cos\left(\frac{\pi(i-k)}{N}\right)+(-1)^{\frac{i+l-j-k}{N}}\cos\left(\frac{\pi(i-l)}{N}\right)\right]}{\sin\left(\frac{\pi(i-j)}{N}\right)\sin\left(\frac{\pi(i-k)}{N}\right)} \right.
\nonumber\\
 +\frac{J_{ij}J_{jk}J_{jl}\left[2\cos\left(\frac{\pi(i-j)}{N}\right)\cos\left(\frac{\pi(j-l)}{N}\right)+(-1)^{\frac{i+l-j-k}{N}}\cos\left(\frac{\pi(j-k)}{N}\right)\right]}{\sin\left(\frac{\pi(j-i)}{N}\right)\sin\left(\frac{\pi(j-l)}{N}\right)} 
\nonumber\\
 +\frac{J_{ik}J_{jk}J_{kl}\left[2\cos\left(\frac{\pi(i-k)}{N}\right)\cos\left(\frac{\pi(k-l)}{N}\right)+(-1)^{\frac{i+l-j-k}{N}}\cos\left(\frac{\pi(j-k)}{N}\right)\right]}{\sin\left(\frac{\pi(k-i)}{N}\right)\sin\left(\frac{\pi(k-l)}{N}\right)} 
\nonumber\\
 \left. +\frac{J_{il}J_{kl}J_{il}\left[2\cos\left(\frac{\pi(l-j)}{N}\right)\cos\left(\frac{\pi(l-k)}{N}\right)+(-1)^{\frac{i+l-j-k}{N}}\cos\left(\frac{\pi(i-l)}{N}\right)\right]}{\sin\left(\frac{\pi(l-j)}{N}\right)\sin\left(\frac{\pi(l-k)}{N}\right)} \right).
\label{eq:V4intper}
\end{eqnarray}
\end{widetext}

For the parent $XY$ model with nonzero nearest and next neighbor interactions the generated diagonal interaction has the slightly modified form compared to  Eq. (\ref{eq:3rdDiagMinim})
\begin{eqnarray}
U_{ij}=2\Delta\left(\delta_{i,j-1}-\delta_{i, j-2}\right), ~ i<j, 
\nonumber\\
\Delta=\eta(N)\frac{J_{1}^2J_{2}}{4F^2}, ~
\eta(N)=\pi^2\frac{2\cos\left(\frac{2\pi}{N}\right)+1}{3N^2\sin\left(\frac{\pi}{N}\right)^2}. 
\label{eq:3rdDiagMinimPer}
\end{eqnarray}
The same constant $\Delta$ determines the four spin hopping amplitude in Eq. (\ref{eq:EffHOffD-resper}) for four consecutive neighboring spins, while it is zero for all other spin quartets. Since the factor $\eta(N)$ very quickly approaches unity with increasing $N$ (for instance $\eta(15)=0.9563$ for the minimum number of spins studied experimentally in Ref. \cite{Monroe2021observationStarkMBL}) we ignore its difference from unity 
and consider the minimalist model in the form of Eq. (\ref{eq:MinimHamCons}). 

{\it The spatial periodicity  can be realized similarly in any other system  with large field gradient } including e. g. Ref. \cite{Bloch21StarkMBL}.

\subsection{Relevance of the Minimalist Model}
\label{sec:ModRel}

Since the minimalist model in Eq. (\ref{eq:MinimHamCons}) is derived as the expansion of the effective Hamiltonian in inverse field gradient $F$ this model should be relevant at sufficient large field gradient $F>F_{c}$. Here we  summarize the estimates of  the crossover field gradient $F_{c}$. while the details are given  in Appendix \ref{sec:AppRelev}. 

The relevance of the minimalist model can be examined conservatively requiring the weak modification of eigenstate energies or liberally requiring the weak change in observables. In the present paper we consider an imbalance in the infinite time limit as the observable characterizing the system dynamics (see Sec. \ref{sec:GrImb} and Appendix \ref{sec:Imbal}).  In Appendix \ref{sec:AppRelev}  energy levels and imbalances were compared for the minimalist model and more accurate models. All comparisons are performed for the $XY$ model in Eq. (\ref{eq:PerH}) with nearest and next neighbor interactions different from zero and identical similarly to the experimentally investigated system in Ref. \cite{Guo2020MBLTiltedModel}.  We set both interactions equal to $J$. 

The conservative  estimate can be obtained considering the minimalist model as the zeroth order  Hamiltonian and the fourth order correction to it  as a perturbation.  Then  a typical perturbation matrix element scales as $V_{4}\sim 2^{-N/2}J^4/F^3$ \cite{Huse15Bath}, while the typical interlevel spacing of the minimalist model scales as $\delta E \sim 2^{-N}J^3/F^2$. Setting $V_{4} \approx \delta E$ we end up with the desirable estimate  that is consistent with that of Appendix II 
\begin{eqnarray}
F_{c1} \approx 0.25\cdot 2^{N/2}J,
\label{eq:FcEstM}
\end{eqnarray} 
obtained comparing the exact diagonalization results for the minimalist model and its corrected version. 

The liberal estimate can be derived requiring the Schrieffer Wolff expansion to be generally converging. Since the expansion parameter of the effective Ha,miltonian is $J/F$  the liberal criterion reads 
\begin{eqnarray}
F_{c2} \approx  J,
\label{eq:FcEstMLib}
\end{eqnarray} 

Even the conservative estimate can be insufficient since an arbitrary small non-local interaction emerging in higher orders in $F^{-1}$ can destroy shuttering and, consequently, localization. Our numerical analysis of imbalance shows that it does not happen.  Moreover, the analysis of imbalances reported in Appendix II shows that the condition Eq. (\ref{eq:FcEstMLib}) is nearly sufficient for the qualitative relevance of the imba;ance behavior obtained in the minimalist model, while the quantitative relevance of this model for localized states requires Eq. (\ref{eq:FcEstM}) to be satisfied. The liberal estimate is relevant quantitatively for delocalized groups of states. 

The present conclusions for the liberal criterion are based on the numerical analysis limited to a relatively small numbers of spins $N \leq 16$. We hope that its predictions can be extended to larger sizes; yet this is the subject for more accurate theoretical or experimental verifications.

\section{Groups of states: localization and delocalization within the minimalist model.}
\label{sec:LocDeloc}

\subsection{Inverted representation and pair hopping}
\label{sec:GrInv}

The spin hopping in the minimalist model is  represented by simultaneous hopping of two  neighboring spins in opposite directions as $\uparrow\downarrow\downarrow\uparrow \leftrightarrow   \downarrow\uparrow\uparrow\downarrow$ in consecutive spin quartets with oppositely oriented middle and border spins. In this picture the spin hopping is hard to trace visually.  However, it is made easier by inverting each second spin as $\uparrow\uparrow\downarrow\downarrow \leftrightarrow   \downarrow\downarrow\uparrow\uparrow$. Then the spin hopping is represented by the hopping of pairs of spins oriented identically. 

The transition from the original model Eq. (\ref{eq:MinimHamCons}) to the inverted spin chain representation can be performed rotating all even numbered spins about the $x$ axis by the angle $\pi$. The transformation unitary matrix for this rotation is given by $\widehat{U}=\prod_{k=1}^{N/2}e^{i\pi S_{2k}^{x}}$. It does not change spin operators in odd numbered sites modifying operators at even numbered sites as $S_{2k}^{x} \rightarrow \widehat{U}S_{2k}^{x} \widehat{U}^{-1}=S_{2k}^{x}$, $S_{2k}^{y} \rightarrow \widehat{U}S_{2k}^{y} \widehat{U}^{-1}=-S_{2k}^{y}$, $S_{2k}^{z} \rightarrow \widehat{U}S_{2k}^{z} \widehat{U}^{-1}=-S_{2k}^{z}$. Consequently $S^{+}$ and $S^{-}$ operators are interchanged at even numbered sites, i. e.   $S_{2k}^{+} \rightarrow S_{2k}^{-}$ and $S_{2k}^{-} \rightarrow S_{2k}^{+}$. This modifies the minimalist model Hamiltonian Eq. (\ref{eq:MinimHamCons}) as 
\begin{eqnarray}
\widehat{H}_{min}=\Delta\sum_{k=1}^{N}(S_{k}^{+}S_{k+1,p}^{+}S_{k+2,p}^{-}S_{k+3,p}^{-}+H.C.)
\nonumber\\
 - 2\Delta\sum_{k=1}^{N}S_{k}^{z}(S_{k+1,p}^{z}+S_{k+2,p}^{z}),
\label{eq:MinimHamConsInv}
\end{eqnarray}
where $S_{k,p}^{a}=S_{k}^{a}$ for $k\leq N$ and  $S_{k,p}^{a}=S_{k-N}^{a}$ for $k> N$ ($a=+$,  $-$ or $z$).  This transformation is applicable only to states containing  an even number of spins  considered below, while the generalization of the results to an odd number of spins is given later in Sec. \ref{sec:GrOdd}.

Inverted chain representation makes spin hopping truly visible. Indeed, consider the inverted state $A$ composed by two sequences of oppositely oriented adjacent spins containing odd numbers of spins $\uparrow\uparrow\uparrow\uparrow\uparrow\downarrow\downarrow\downarrow\downarrow\downarrow\downarrow\downarrow$. 
We will refer to such sequences as {\it odd} sequences, while sequences containing even numbers of spins are referred as {\it even} sequences. 
It is clear from this picture how spin pairs can propagate through the chain. For example, the rightmost  upwards oriented pair of fourth and fifth spins can hop   three times to the right reaching tenth and eleventh positions (state $B$: $\uparrow\uparrow\uparrow\downarrow\downarrow\downarrow\downarrow\downarrow\downarrow\uparrow\uparrow\downarrow$), while the leftmost  pair of downwards oriented spins (sixth and seventh spins) can hop to the left maximum two times reaching second and third positions (state $C$: $\uparrow\downarrow\downarrow\uparrow\uparrow\uparrow\uparrow\downarrow\downarrow\downarrow\downarrow\downarrow$). These states $A$, $B$ and $C$ for the original spin chain look like  $\uparrow\downarrow\uparrow\downarrow\uparrow\uparrow\downarrow\uparrow\downarrow\uparrow\downarrow\uparrow$, 
$\uparrow\downarrow\uparrow\uparrow\downarrow\uparrow\downarrow\uparrow\downarrow\downarrow\uparrow\uparrow$ and  $\uparrow\uparrow\downarrow\downarrow\uparrow\downarrow\uparrow\uparrow\downarrow\uparrow\downarrow\uparrow$, respectively. It is much more difficult to establish the visual connection between the latter three states compared to these states for the inverted chain.  

The inverted chain representation permits us to characterize spin hopping using the parity of sequences of adjacent spins having identical orientations. Odd or even sequences are sequences containing odd or even number of identically oriented consecutive spins, respectively. For instance the state  $\uparrow\uparrow\uparrow\downarrow\downarrow\downarrow\downarrow\downarrow\downarrow\uparrow\uparrow\downarrow$ is composed by sequences of $3$, $6$, $2$ and $1$ spins. One can define it by the set of numbers $\{3, 6, 2, 1\}$ specifying that the first spin of the first sequence is located in the first position of the periodic chain and it is oriented upwards. Due to the periodicity of the chain  the state $\uparrow\uparrow\downarrow\downarrow\downarrow\downarrow\downarrow\downarrow\uparrow\uparrow\downarrow\uparrow$ is defined by the same set of numbers with the second spin of the first sequence  located in the first position of the periodic chain and oriented upwards.

\subsection{Classification of Krylov subspaces: localized and delocalized groups of states. }
\label{sec:GrGr}

Product states coupled by spin pair hops form the basis for system eigenstates and this basis defines the Krylov subspace for a specific group of states  \cite{Moudgalya2019thermalization,Bardarson21FragmentedSpaceSampleSt}. 
The  basis  {\it product} states of an {\it inverted} chain can be represented as the set of consecutive numbers of identically oriented spins   $\{n_{1},n_{2},...n_{p}\}$ (e. g.  $\{2,3,4,3\}$ for the state $A$ in Eq. (\ref{eq:Constr1})). If sequences $1$ and $p$ are oriented identically, then they should be considered jointly as $\{n_{1}+n_{p},n_{2},...n_{p-1}\}$ with the specified position of the first spin. 
Pair hopping conserves the number of odd sequences since it can modify the sequence size only by $2$. Therefore the numbers of odd sequences are identical in all product states belonging to the given Krylov subspace, serving  conserved quantities  (cf.  Ref. \cite{Rakovszky2020}).

{\it The main results of the present work} can be formulated in terms of the relative parities of sequences shared between all product states belonging to the given Krylov subspace as following. If all odd- or even-numbered sequences are even (except for frozen states possessing  all identical spins or having a single spin in all even- or odd-numbered sequences) then corresponding Krylov subspaces abd system eigenstates possess  the translational invariance with the period $2$ (see e. g. states in Eq.  (\ref{eq:Constr1})). Consequently,, these states are delocalized. The remaining  product states having odd sequences at both odd and even positions form Krylov subspaces  (with marginal exceptions) with no translational invariance and confined spin transport. 

Product states corresponding to translationally invariant Krylov subspaces can be separated into two  groups of states enumerated by the Roman numbers {\bf I} and  {\bf II}. The  group {\bf I}  consists of $2\cdot 2^{N/2}$  states composed by all even sequences. The group {\bf I} states  can be mapped to $N/2$ spins $1/2$ representing pairs  \cite{Bergholtz05QHallXYmodel} (spin subspaces of 
Refs.   \cite{Moudgalya2019thermalization,Bardarson21FragmentedSpaceSampleSt}).  

\begin{figure}
\includegraphics[scale=0.4]{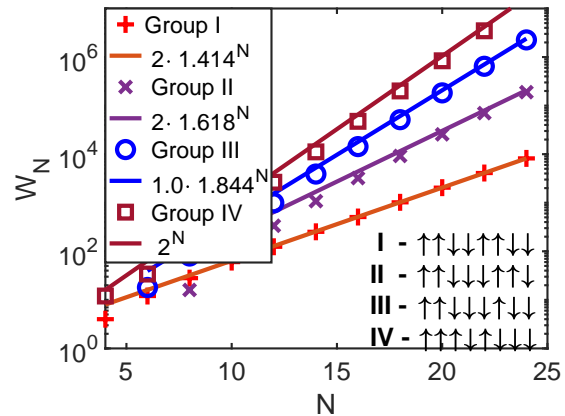} 
\caption{ Numbers of states  vs the numbers of spins for all groups  with representative states for all groups.}
\label{fig:Nums}
\end{figure}

The states of the group {\bf II} are formed by the product states 
possessing at least one odd sequence and all even  sequences of spins oriented either upwards or downwards as in Eq.  (\ref{eq:Constr1}).  Correspondingly, the odd sequences must have the  opposite orientation. 
The number of states belonging to  this group {\bf II} behaves as  $W_N\approx 2\cdot ((1+\sqrt{5})/2)^{N} \propto 1.618^N$ at large $N$.  The numbers  of states are calculated for all groups in 
Appendix  \ref{sec:GenerFunc}, using the generating function method, and presented in Fig. \ref{fig:Nums} together with the representative group states. All sectors of the Hilbert space spanned by states lying in group {\bf I}, correspond to integrable models  \cite{Bardarson21FragmentedSpaceSampleSt},  while the  group  {\bf II} states are mostly ergodic as pointed out  below in Sec. \ref{sec:GrDeriv}.  

The states of both groups must have even numbers of spins. The original (non-inverted) states are characterized by a zero spin projection to the $z$ axis. The zero spin projection corresponds to the filling factor $1/2$ leading to at least two fold degeneracy of all states \cite{SeidelQHall05}, which is realized for eigenstates of the problem Eq. (\ref{eq:MinimHamCons}) shifted by the one lattice period with respect to each other, since they cannot be coupled by pair hopping.  

  
The remaining non-translationally invariant states can be represented by states containing at least two  odd sequences with oppositely oriented spins. We separate  them  into two groups including the group  {\bf III} of states having  no frozen spins, with conserved projections to the $z$-axis,   and group {\bf IV} of states containing frozen  spins as defined in Eq. (\ref{eq:Constr}) (for example, third, fourth and fifth spins are frozen  in the representative state of the group {\bf IV} in Fig. \ref{fig:Nums}). The  group {\bf IV} contains the majority of states ($\propto 2^{N}$)  including $1.134\cdot 1.7549^N$ completely frozen states with all frozen spins, see  Appendix  \ref{sec:GenerFunc}. 
Localization obviously takes place for the group {\bf IV} states split into independent blocks by frozen spins.  The group {\bf III} states are mostly localized for  the present model Eq. (\ref{eq:MinimHamCons}) in accord with the imbalance behavior  (see Fig. \ref{fig:Imb}). 

The states with odd number of spins   belong either to the group {\bf III} or {\bf IV} depending on the presence  of  frozen  spins Eq. (\ref{eq:Constr}) (see Sec. \ref{sec:OddNums}).


\subsection{Translational invariance. }
\label{sec:GrDeriv}
 
Here we prove  the translational invariance for the groups  {\bf I} and {\bf II} and its absence for almost all Krylov's subspaces in groups {\bf III} and {\bf IV}.  To prove the translational invariance for the  groups {\bf I} or {\bf II}   consider the product state belonging to these  groups with even number of spins in all odd-numbered sequences. Such state is illustrated by the 
 state  $A=\{2,3,4,3\}$ defined below as 
\begin{eqnarray}
({\rm \mathbf{A}}) \overset{1}{\SpinUp}\overset{2}{\SpinUp}\overset{3}{\SpinDown}\overset{4}{\SpinDown}\overset{5}{\SpinDown}\overset{6}{\SpinUp}\overset{7}{\SpinUp}\overset{8}{\SpinUp}\overset{9}{\SpinUp}\overset{10}{\SpinDown}\overset{11}{\SpinDown}\overset{12}{\SpinDown}  \rightarrow  ({\rm \mathbf{B}}) 
\overset{1}{\SpinUp}\overset{2}{\SpinUp}\overset{3}{\SpinDown}\overset{4}{\SpinDown}\overset{5}{\SpinDown}\overset{6}{\SpinUp}\overset{7}{\SpinUp}\overset{10}{\SpinDown}\overset{11}{\SpinDown}\overset{8}{\SpinUp}\overset{9}{\SpinUp}\overset{12}{\SpinDown} \rightarrow
\nonumber\\
({\rm \mathbf{C}})\overset{1}{\SpinUp}\overset{2}{\SpinUp}\overset{3}{\SpinDown}\overset{4}{\SpinDown}\overset{5}{\SpinDown}\overset{10}{\SpinDown}\overset{11}{\SpinDown}\overset{6}{\SpinUp}\overset{7}{\SpinUp}\overset{8}{\SpinUp}\overset{9}{\SpinUp}\overset{12}{\SpinDown}  \rightarrow  ({\rm \mathbf{D}}) 
\overset{3}{\SpinDown}\overset{4}{\SpinDown}\overset{1}{\SpinUp}\overset{2}{\SpinUp}\overset{5}{\SpinDown}\overset{10}{\SpinDown}\overset{11}{\SpinDown}\overset{6}{\SpinUp}\overset{7}{\SpinUp}\overset{8}{\SpinUp}\overset{9}{\SpinUp}\overset{12}{\SpinDown} ~~~~\hspace{0.05cm}
\label{eq:Constr1}
\end{eqnarray} 
The translational invariance with the period $2$ for these states can be demonstrated  considering  spin pair hopping starting with the rightmost sequence $k$ in even numbered position having more than one spin ($n_{k}>1$) (spins $10$, $11$, $12$ in the state $A$  in Eq. (\ref{eq:Constr1})). If there is only one spin there then another even numbered sequence $k$ should be considered with $n_{k}>1$.  One can take the leftmost pair of spins in this sequence and move it to the left until joining the left next neighboring sequence $k-2$ that is possible because $k-1^{st}$ sequence is even (transition  $A\rightarrow B\rightarrow  C$  in Eq. (\ref{eq:Constr1})). Then the leftmost  spin pair of $k-2^{nd}$ sequence (spins $3$, $4$)  moves left to join $k-4^{th}$ sequence ($C\rightarrow D$). This procedure should be continued until the pair of spins will join the $k^{th}$ sequence from the right (the state $D$  in Eq. (\ref{eq:Constr1})). The final state ($D$) is formed  by the translation of the initial state ($A$)  to the right by two steps. This  proves the translational invariance of the associated Krylov subspace and, consequently, eigenstates of the problem. 

Using the similar arguments one can give a full description of Krylov subspaces of groups {\bf I} and {\bf II}. The states of the group {\bf I} for $N$ spins and total spin projection $S$ to the $z$ axis for the inverted states  (remember that for the group {\bf I} $N$ must be even and $S$ must be integer) belong to two Krylov subspaces of states made of even sequences with the sequence borders located all either in even or odd positions.

The states and Krylov's subspaces of the group {\bf II}  can be further characterized using the analysis similar to that in Ref. \cite{Rakovszky2020} for  non-local integrals of motion. Such analysis is beyond the scope of the present paper targeted to distinguish delocalized, translationally invariant states of groups {\bf I} and {\bf II} and almost all localized states of groups 
{\bf III} and {\bf IV}.

Krylov subspaces containing states with both odd- and even-numbered  oppositely oriented  odd sequences have no translational invariance except for marginal situations including, for instance, Krylov subspaces containing periodic  states (e. g.  $\{3,3,3,3\}$). 

To prove that consider the closest oppositely oriented odd sequences in odd and even positions. They are separated by the even number of even sequences. These sequences can be removed by means of pair hops from each sequence towards the closest odd sequences with the same direction of spins. Since    the spin pair located between two odd sequences cannot hop through them the position of this border is conserved within the Krylov subspace. 

The position of the boundary between two odd sequences in that configuration is unique and it cannot be modified by the pair hopping because of the spin projection conservation to the $z$ axis. Odd sequences are not transparent for pair hopping because pairs move by two steps only. Therefore adjacent odd sequences confine the  spin transport breaking down the translational invariance of corresponding Krylov subspaces.  This is not true for the group {\bf II} states where  all odd sequences possess  the same spin orientation.  Indeed, they can exchange by pairs that can pass through even sequences separating them. 

The fixed position of the boundary between neighboring  odd sequences in the product state with the minimized  number of sequences violates the translational invariance of the Krylov subspace with the only exception of subspaces containing the translationally invariant states composed by self-repeating sets of spins.  The simplest translationally invariant states belonging to the group {\bf III} are given   by self repeating sequences   $\{2p+1, 2p+1,...2p+1\}$ with any integer $p$  
 \cite{Moudgalya20QHall}. 


\subsection{Frozen spins} 
\label{sec:GrFrz}
 
The majority of states  of the system ($\sim 2^{N}$) belong to the shattered group {\bf IV} because of the existence of a finite length frozen spin groups   \cite{Pollmann20Fragment,Bardarson21FragmentedSpaceSampleSt} in the large number of spins limit. Indeed, if there exist the frozen spin group of a finite length it can be characterized by the formation probability $P_{f}$ per the unit chain length. Consequently, in the large number of spins limit the probability to avoid that group scales approximately as $e^{-P_{f}N}$ suggesting exponentially small weight of states lacking frozen spins, 
   
How are these frozen groups formed? If a sequence is even and at least one of its neighboring sequences possesses more than one spin, then all spins within that even sequence are mobile since the pair of spins from the neighboring sequence can hop through the sequence under consideration shifting all its spins by two (see e. g. $A \rightarrow C$ transition in Eq. (\ref{eq:Constr1})). Therefore an even sequence can belong to the frozen set only if its neighboring sequences contain  
only one spin.  

If an odd sequence has both neighbors containing more than one spin it is mobile. However, if one of its neighbors   has the only one spin then its edge spin on that side cannot hop  until that neighboring sequence  changes. Consequently, a single spin sequence surrounded by two odd sequences (see case A  below)
\begin{eqnarray}
({\rm \mathbf{A}})~ \overset{1}{\SpinUp}\overset{2}{\SpinUp}\overset{3}{\SpinDown}\overset{4}{\SpinDown}\boxed{\overset{5}{\SpinDown}\overset{6}{\SpinUp}\overset{7}{\SpinDown}}\overset{8}{\SpinDown}\overset{9}{\SpinDown}\overset{10}{\SpinDown}\overset{11}{\SpinDown}\overset{12}{\SpinUp}  \hspace{1.25cm}
\nonumber\\
({\rm \mathbf{B}})~ \SpinUp\SpinUp\SpinDown\SpinDown\boxed{\SpinDown\SpinUp\left[\SpinDown\underset{\rm \bf even}{\bullet\bullet\bullet}\SpinDown\SpinUp\right]_{n}\SpinDown}\SpinDown\SpinDown\SpinDown\SpinDown\SpinUp
\label{eq:Constr}
\end{eqnarray} 
forms a simplest frozen  set $\{odd, 1, odd\}$ with three frozen spins at positions $(5,  6,  7)$ shown  within the box in the state $A$ in Eq. (\ref{eq:Constr}).   If the set $\{odd, 1\}$ is followed by the sequence with an even number of spins it should have the next sequence containing only a single spin to keep spins being frozen. The fragment $\{even, 1\}$ can be added an arbitrary number of times ($n$ in Eq. (\ref{eq:Constr}), state B) until being terminated by the odd sequence. This is the only  way of creating a finite size frozen set of spins just by construction. 

Completely frozen states must have each sequence with more than one spin being surrounded by single spin sequences (except for the state of all identical spins).  The number of such states belonging to the group {\bf IV} increases with the number of spins  as $1.134\cdot 1.7549^N$, see Appendix  \ref{sec:GenerFunc}.  

\subsection{States possessing odd number of spins}
\label{sec:OddNums}
\label{sec:GrOdd}

 We cannot invert spins in a periodic chain with an odd number of spins  since odd positions become even after passing the period as shown below: 
\begin{eqnarray}
 (\mathbf{A}) \downarrow\uparrow\downarrow\downarrow\uparrow ~\rightarrow ~ (\mathbf{A}_{1})  \downarrow\downarrow\uparrow\uparrow\downarrow, \hspace{0.8cm}
\nonumber\\
 (\mathbf{B})  \overset{1}{\downarrow}\overset{2}{\downarrow}\overset{3}{\downarrow}\overset{4}{\uparrow}\overset{5}{\uparrow}~\rightarrow ~ (\mathbf{B}_{1}) \overset{1}{\downarrow}\overset{4}{\uparrow}\overset{5}{\uparrow}\overset{2}{\downarrow}\overset{3}{\downarrow} ~ \xcancel{\rightarrow} ~ (\mathbf{B}_{2})\overset{5}{\uparrow}\overset{3}{\downarrow} \overset{1}{\downarrow}\overset{2}{\downarrow}\overset{4}{\uparrow}, \hspace{-0.5cm}
\nonumber\\
 (\mathbf{C})  \overset{1}{\downarrow}\overset{2}{\downarrow}\overset{3}{\downarrow}\overset{4}{\uparrow}\overset{5}{\uparrow}\overset{6}{\uparrow}\overset{7}{\uparrow}\overset{8}{\uparrow}\overset{9}{\downarrow}\overset{10}{\downarrow}~ \rightarrow  ~ (\mathbf{C}_{1}) \overset{1}{\downarrow}\overset{4}{\uparrow}\overset{5}{\uparrow}\overset{2}{\downarrow}\overset{3}{\downarrow}\overset{6}{\uparrow}\overset{9}{\downarrow}\overset{10}{\downarrow}\overset{7}{\uparrow}\overset{8}{\uparrow}. \hspace{-0.4cm}
\label{eq:Odd}
\end{eqnarray} 
 In the inverted chain with overturned spins in even positions   the state $A$ transforms to  the state $B$.  According  to the rules for the pair hopping  in addition to  the pair hopping leading to the state $B_{1}$  (the straight chain state $A_{1}$) there is the pair hopping through the border to the state $B_{2}$, which is not permitted for the straight state. This problem can be resolved adding the second chain that is the fully inverted copy of the first chain (see the state $C$ in Eq. (\ref{eq:Odd})). Then the pair hopping should be performed simultaneously for the pair and its copy like it is shown for the transition $C\rightarrow C_{1}$ where the hopping of the pair of spins $2$ and $3$ by two steps to the right is accompanied by the hopping of its inverted copy  (spins $(7, 8)$) by two steps to the right. There is only one allowed pair hopping from the state $C_{1}$ backwards to the state $C$. Thus this double chain with partially inverted spins have hopping of pairs equivalent to dipole moment conserving transitions in the straight spin chain.  

Consequently, one can construct the Krylov subspace in double inverted chain using  simultaneous   hopping of  pairs and their copies shifted by $N$. 
Similarly to the previous  consideration a simultaneous existence of odd sequences in even and odd positions  breaks down the translational invariance of the Krylov subspace.  It turns out that such sequences exist inevitably in any state of $2N$ spins for an odd number of spins $N$. 

Indeed, for an odd total  number of spins $N$  at least one sequence containing an odd number of spins must exist. Then its copy with the opposite spin orientation must exist as well in the copy state. Consequently, we have two sequences with odd numbers of spins occupying odd and even positions (because they are of opposite orientations) that contradicts to the requirement for all either odd- or even-numbered sequences to have even numbers of spins, which is the necessary requirement for the translational invariance.

Thus we show that spin states having an odd numbers of spins always have confined spin transport and belong either to the group {\bf III} if all spins are mobile or the group {\bf IV} if it contains  frozen spins Eq. (\ref{eq:Constr}). 

\begin{figure}
\includegraphics[scale=0.45]{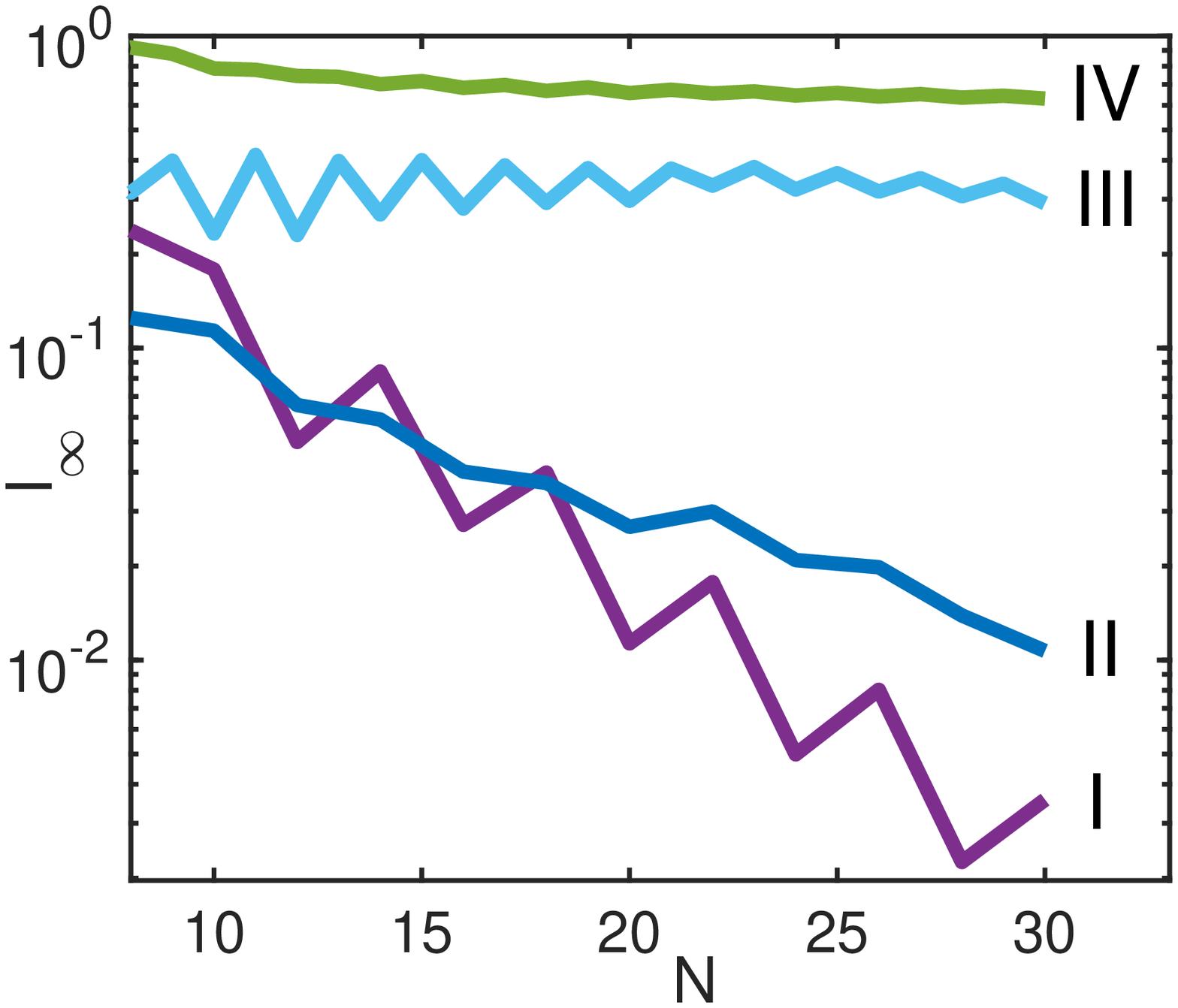}
\caption{Dependence of average imbalances evaluated within the infinite time limit on the number of spins for different groups  (solid lines with the group numbers in the right).}
\label{fig:Imb}
\end{figure}

\subsection{Group averaged imbalances}
\label{sec:GrImb}

Here we consider the connection between the  groups, defined in Sec. \ref{sec:GrGr},  
and dynamic properties of states belonging to these groups expressed in terms of experimentally accessible imbalance determined by the evolution of the initial product state \cite{Bloch16MBLExp,Monroe2021observationStarkMBL}.  
The average imbalance for the initial product state  $a$ is defined as \cite{Monroe2021observationStarkMBL}  
\begin{eqnarray}
I_{a}(t)=\frac{\frac{1}{N}\sum_{k=1}^{N}<S_{k}^{z}(t)><S_{k}^{z}(0)>-<S_{k}^{z}>_{*}^2}{1/4-<S_{k}^{z}>_{*}^2}.
\label{eq:ImbDefMT}
\end{eqnarray}
where $<S_{k}^{z}(0)>$ is the projection of the spin $k$ to the $z$ axis in the  state $a$,  $<S_{k}^{z}(t)>$ is the average projection of this spin to the $z$ axis at time $t$ and $<S_k^{z}>_{*}$ is the expectation value for the  projection of the spin to the $z$ axis in the ergodic system. With this definition the imbalance should approach zero in the ergodic regime at an infinite time and in the thermodynamic limit of an infinite number $N$ of spins \cite{Srednicki94ETH}.  
In the localized regime it should  remain constant. The convergence of imbalance to its expectation value in the  ergodic, delocalized regime has to be exponential as $N$ approaches infinity. 

This criterion is used below to distinguish  localized and delocalized states. The delocalized regime is not necessarily ergodic \cite{Moudgalya2019thermalization} and therefore we discuss the level statistics  in the end of the present section. 

How to define correctly expectation value $<S_{k}^{z}>_{*}$ in the ergodic regime? In the infinite temperature limit we  assume that all states contributing to the eigenstates  should be represented equally in this average. Consequently, spin projections should be averaged over the Krylov subspace defined  for each given initial state. 

\begin{figure}
\includegraphics[scale=0.45]{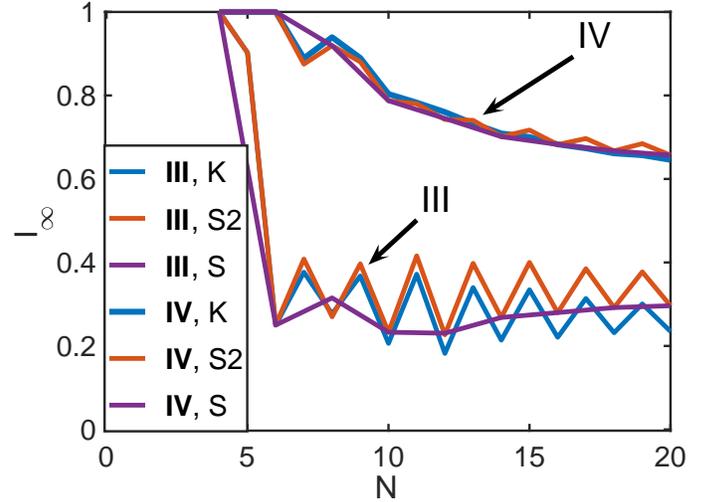}
\caption{Comparison of average imbalances for groups  {\bf III} and {\bf IV} with expectation values evaluated for different subspaces. Letter  $K$ means Krylov subspace averaging,  letters $S2$ means averaging over subspaces with fixed odd and even sublattice spin projections to the $z$ axis for even number of spins, while $S$ means averaging with the  fixed total spin projection to the $z$ axis.}
\label{fig:ImbSubsp1}
\end{figure}

It is straightforward to define average spin projections for Krylov subspaces corresponding to the translationally invariant groups {\bf I} and {\bf II} for even numbers $Н$ of spins. The Hamiltonian Eq. (\ref{eq:MinimHamCons}) conserves the total projections of odd and even numbered spins to the $z$ axis ($S_{odd}=\sum_{k=1}^{N/2}S_{2k-1}^z$ or $S_{even}=\sum_{k=1}^{N/2}S_{2k}^z$, respectively). Then the Krylov subspace averaged projection of spin $k$ to the $z$ axis is given by $2S_{odd}/N$ or $2S_{even}/N$ for odd or even $k$, respectively, due to the translational invaariance of the Krylov subspace with the period $2$.  We used this definition of expectation values in Eq. (\ref{eq:ImbDefMT}) for even number of spins and arbitrarily group of the initial state. For odd number of spins we set the expectation value for the average spin projection equal to its average value $\sum_{k=1}^{N}S_{k}^z/N$.  With this definition of the spin projection expectations we evaluated group averaged imbalances in the infinite time limit as shown in Fig. \ref{fig:Imb}. The Krylov subspace averaged imbalances for groups {\bf III} and {\bf IV}  are quite close to those evaluated with spin projection expectation values used in Fig. \ref{fig:Imb}  as illustrated in Fig. \ref{fig:ImbSubsp1}. Therefore our definition of expectation values is relevant for all states that is important for the analysis of experimental data.  Indeed,  it is streightforward to evaluate spin projections using the initial states, while the evaluation of Krylov subspace averaged projections can be problematic because of the exponentially large number of states. 

The group averaged imbalances represented in Fig. \ref{fig:Imb} are obtained averaging imbalances for specific state $a$ over all states belonging to a certain group.  They are given in an infinite time limit ($I_{\infty}$)  \cite{BerezinskiiGorkov79}. 
The infinite time limit of  imbalance is  evaluated numerically expanding the time dependent system  wavefunction over the basis of  eigenstates $|\alpha>$ with eigenenergies $E_{\alpha}$, obtained using exact diagonalization of the system Hamiltonian,   as (remember that we set $\hbar=1$)
\begin{eqnarray}
|\psi(t)>=\sum_{\alpha} <\alpha |a> |\alpha>e^{-iE_{\alpha}t}.  
\label{eq:EigStwfexp}
\end{eqnarray}

Consequently, the average spin projection at the time $t$ ($<S_{k}^{z}(t)>$) can be expressed as $\sum_{\alpha,\beta} e^{-i(E_{\alpha}-E_{\beta})t}<\alpha |a><a |\beta><\beta|S_{k}^{z}|\alpha>$. In the infinite time limit we leave only terms with $E_{\alpha}=E_{\beta}$, while oscillating terms are averaged out.  This yields 
\begin{eqnarray}
<S_{k}^{z}(\infty)>
\nonumber\\
=\sum_{\alpha,\beta} <\alpha |a><a |\beta><\beta|S_{k}^{z}|\alpha>\delta_{E_{\alpha},E_{\beta}}, 
\label{eq:ImbInfTime}
\end{eqnarray}
where the generalized Kronecker symbol  $\delta_{E_{\alpha},E_{\beta}}$ is equal to unity for $E_{\alpha}=E_{\beta}$ and $0$ otherwise. 

The infinite time limit is accessible experimentally using time averaged imbalance as demonstrated in  Appendix \ref{sec:Imbal}.

The dependence  of group average imbalances on the number of spins is consistent with our expectations. The average  imbalances approach zero  exponentially with increasing   $N$ for delocalized states of groups {\bf I} and {\bf II} and remains finite for other two  groups with increasing $N$. 

The convergence of imbalance to its expectation value in Fig. \ref{fig:Imb} is exponential in the number of spins $N$ for groups {\bf I} and {\bf II} as it is expected for the ergodic behavior. However, the additional analysis of the level statistics (to be reported elsewhere) suggests ergodic behavior only for the states belonging to the group {\bf II}, that is consistent with earlier expectations for the Bethe anzatz integrable nature of the group {\bf I} states for the present problem Eq. (\ref{eq:MinimHamCons}). However, the addition of a small amount of quenched disorder should make the system ergodic \cite{Kudo2004LevStatBethAnzDis}.

The actual imbalance is affected by the Schrieffer-Wolff rotation modifying the definitions of the spin projection operators compared to the minimalist model Fig. \ref{fig:ImbSubsp1}. However, we expect this effect of order of $(J/F)^2$ to be of a minor significance already at $F>3J$.  

\section{Discussion of the recent experiments  \cite{Guo2020MBLTiltedModel,Monroe2021observationStarkMBL}: How to observe delocalization at large field gradients?}
\label{sec:Exp}

Here we discuss  the spin systems  investigated experimentally  in Refs. \cite{Guo2020MBLTiltedModel,
Monroe2021observationStarkMBL}  that are similar to those considered in the present work. For a large field  gradient $F \approx  2.5J$ a substantial localization was observed in these experiments for all probed initial states in contrast with the expected coexistence of localized and delocalized states.  In our opinion this is the consequence of system inhomogeneity due to the open boundaries \cite{Guo2020MBLTiltedModel,
Monroe2021observationStarkMBL} and the 
lack of some spin-spin next neighbor interactions in the system, investigated in  Ref.  \cite{Guo2020MBLTiltedModel}.
Below we show for both systems that removing inhomogeneity and making them periodic  following the receipt of Sec. \ref{sec:ModelPBC}   one can observe delocalization of  states belonging to the groups  {\bf I} and {\bf II} at arbitrary large field gradient.

\subsection{Interacting qubits within the superconducting quantum processor \cite{Guo2020MBLTiltedModel}.}
\label{sec:ExpQub}

Two systems of $16$ and $29$  qubits within the superconducting quantum processor were  investigated in Ref. \cite{Guo2020MBLTiltedModel}. Both systems can be represented by $XY$ models  of  interacting spins $1/2$ (qubits)  placed into uniformly growing field. For $16$ spins only nearest neighbor interactions were used so the system can be reduced to free fermions  \cite{JordanWigner28}. For this system  the full Wannier -- Stark localization of all states always takes place so we do not consider it.   

\begin{figure}
\includegraphics[scale=0.45]{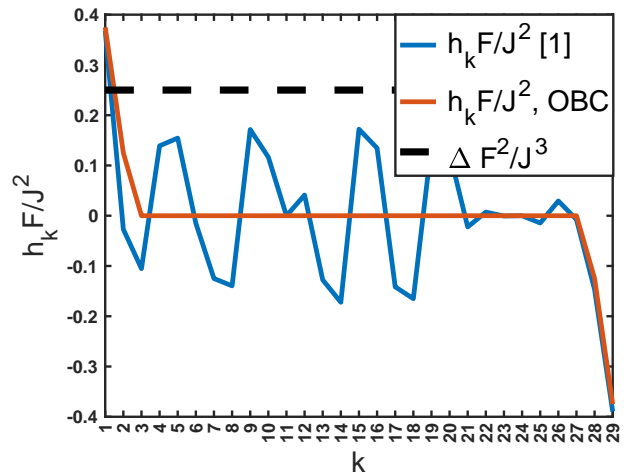} 
\caption{Longitudinal fields vs. the number of spin $k$ generated by means of the Schrieffer-Wolff transformation in the model of Ref. \cite{Guo2020MBLTiltedModel} (blue line) and in the OBC model with all identical interactions between nearest and next neighbor spins (red line)  compared to the four spin hopping interaction conserving dipole moment (dashed black line).}
\label{fig:hs}
\end{figure}

Thus the system of our interest is formed by $29$ qubits (spins $1/2$) arranged in a chain with the nearest and next neighbor hopping interactions  $J_{ij}S_{i}^{+}S_{j}^{-}$ all defined  in  Fig.  S2.A  in Supplementary Materials of Ref. \cite{Guo2020MBLTiltedModel} with almost identical interactions $J_{ij}$  for a majority of nearest and next neighbor interactions  except for several lacking next neighbor interactions.  There is no direct interaction between the ends of the chain so the OBC regime is realized. Consequently, at large field gradients the effective Hamiltonian of the system contains  static longitudinal fields $h_{k}$ acting on spins, which are generated in the second order of the Schrieffer-Wolff expansion   Eq. (\ref{eq:sw2nddiag})  in the form 
\begin{eqnarray}
h_{k}=-\frac{1}{4F}\sum_{l\neq k}\frac{J_{kl}^2}{(k-l)}. 
\label{eq:hk}
\end{eqnarray}
These fields are shown in Fig. \ref{fig:hs} where they are rescaled by the factor $J^2/F$ and compared with four spin hopping amplitude $\Delta$ Eq. (\ref{eq:MinimHamCons}) rescaled by the factor  $J^3/F^2$. Their relationship in  Fig. \ref{fig:hs} corresponds to the case of $J=F$. Even in that case a typical field is comparable with  the hopping strength, while for larger  field gradient $F$ the ratio of generated fields and the hopping amplitude
increases proportionally to $F$. Consequently, in this limit the full many-body localization should be naturally expected in agreement with the experimental results \cite{Guo2020MBLTiltedModel}.

\begin{figure}
\includegraphics[scale=0.45]{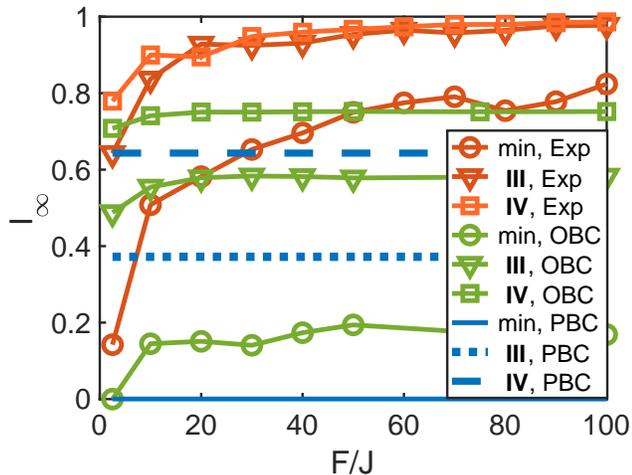} 
\caption{Average infinite time imbalances for groups {\bf III} and {\bf IV} and minimum imbalances vs. a field gradient $F$. The results are given for the models of Ref. \cite{Guo2020MBLTiltedModel} (Exp), OBC and PBC minimalist models.}
\label{fig:imbinfq}
\end{figure}

However, if we set all neighbor and next neighbor interactions equal to each other (leaving other interactions equal to  zero as in Ref. \cite{Guo2020MBLTiltedModel}) then the field is induced only  in the four edge sites of the spin chain, while it is zero in all other sites (see Fig. \ref{fig:hs}). In this case (referred in  Figs. \ref{fig:hs} and   \ref{fig:imbinfq} as the OBC model) delocalization can take place for all spins, except for those at edges, at arbitrarily large field gradient. If in addition the system is made periodic implementing  the time-periodic Hamiltonian of Eq. (\ref{eq:PerH}), then no longitudinal field exists. In the latter case the states should be separable into localized and delocalized groups in accord with Sec. \ref{sec:LocDeloc}. 

Here we report  the investigation of imbalances in all three models referred as that of Ref. \cite{Guo2020MBLTiltedModel}, OBC and PBC within the minimalist models represented by the Schrieffer-Wolff expansion up to the third order in $J/F$ for $F\gg J$. Since the number of spins in Ref. \cite{Guo2020MBLTiltedModel}  is odd ($N=29$) the product  states  can belong only to groups {\bf III} or {\bf IV}.  We evaluated average imbalances within the infinite time limit for these groups for all three models choosing initial state randomly and collecting data until the standard deviation of the average imbalance for each group exceeds $0.5$\%. We also collected minimum imbalances for all considered states to approach most delocalized states. It is natural to expect that the initial state possessing the minimum imbalance belongs to the Krylov subspace  with most delocalized states. Average spin projections  in the initial product states were used for  spin projection expectation values in the definition of imbalance Eq. (\ref{eq:ImbDefMT}) similarly to Sec. \ref{sec:GrImb} for an odd number of spins. 

The Monte-Carlo estimate of imbalances has been compared with  the exact calculations for the system of  $N=20$ spins where it gives quite reasonable estimate both for average and minimum imbalances.  This justifies its application  to $N=29$ spins. 


 Average imbalances for the groups {\bf III} and {\bf IV} and minimum imbalances  are shown in Fig. \ref{fig:imbinfq}. The imbalance behavior indicates substantial localization in the model of Ref. \cite{Guo2020MBLTiltedModel} with increasing the field gradient $F$ where average and minimum imbalances increase with increasing $F$ approaching the maximum value $1$. This is due to longitudinal fields (see Fig. \ref{fig:hs}) suppressing delocalization stronger at larger $F$.  In OBC and PBC models imbalances are almost or completely field gradient independent because there is no generated longitudinal fields for the majority of spins (OBC) or all spins (PBC) in these models. Yet, finite average imbalances suggests substantial localization of states in both models that is consistent with the nature of the groups {\bf III} and {\bf IV}. 
 
 The minimum imbalance for the model generated using experimental parameters \cite{Guo2020MBLTiltedModel} and at  the field gradient $F=2.5J$ is  $I_{\rm min}=0.1423$. It is realized for the initial stare   $\downarrow \uparrow\uparrow   \downarrow \downarrow  \uparrow \downarrow \downarrow\uparrow \uparrow\downarrow \downarrow \uparrow\uparrow \downarrow  \downarrow\uparrow\downarrow\uparrow\downarrow\downarrow\uparrow \uparrow\downarrow\uparrow \downarrow \uparrow \uparrow\downarrow$. The corresponding eigenstates are partially  delocalized, yet representing the minority of states since the average imbalance exceeds $0.6$ (see Fig. \ref{fig:imbinfq}). Further increase of the field gradient leads to the increase of the minimum imbalance (e. g $I_{min}=0.5$ for $F=8J$) indicating the localization of all states due to generated static fields.  
 
 The minimum imbalance for the PBC model $I_{\rm min}=0.27\cdot 10^{-3}$ indicates a substantial delocalization of corresponding states (initial state is $\uparrow \downarrow \downarrow \uparrow \uparrow \downarrow \downarrow \uparrow \downarrow \uparrow \downarrow \uparrow \downarrow \uparrow \uparrow \downarrow \uparrow \downarrow \downarrow \downarrow \uparrow \uparrow \uparrow \downarrow \downarrow \downarrow \uparrow \uparrow \downarrow$). Thus  delocalization  of a minority of states is possible for the group {\bf III} states where all spins are mobile.  
  The minimum imbalance for the OBC model at large field gradient realized for the initial state $\downarrow \downarrow \downarrow \uparrow \downarrow \downarrow \uparrow \uparrow \downarrow \downarrow \uparrow \uparrow \uparrow \downarrow \downarrow \uparrow \uparrow \downarrow \downarrow \uparrow \downarrow \uparrow \uparrow \downarrow \uparrow \downarrow \downarrow \uparrow \uparrow$ is around $0.15$. This estimate is approximately  consistent with the minimum imbalance estimate $4/29 \sim 0.14$ for the fully delocalized state where all spins except for the four spins at the  chain boundaries supposes to have time averages close to zero. Therefore, there can be a  substantial delocalization for some states of the OBC model. For the minimum considered field gradient $F=2.5J$ the minimum imbalance for that model approaches zero. 
 
 Thus delocalization of some states in the system studied in Ref. \cite{Guo2020MBLTiltedModel}  can be attained in the large field gradient limit by making the spin-spin interactions between nearest and next neighbor spins identical. The delocalization can be further strengthened  by means of making the system periodic and using an even numbers of spins as shown in Fig. \ref{fig:Imb} for the groups {\bf I} and {\bf II}.  
 
Our consideration is limited to the minimalist model. The numerical study of more accurate models for the system of $29$ spins is problematic because of the huge number of states involved. Yet we believe that our results remain valid at least qualitatively according to the analysis of Sec. \ref{sec:ModRel}.

\subsection{Trapped ion quantum simulator   \cite{Monroe2021observationStarkMBL}.}
\label{sec:ExpIons}

Here we examine the chain of $N=15$ spins with the hopping interaction $J_{ij}=1/|i-j|^{1.3}$ placed in a uniformly growing field with the gradient $F$, Eq. (\ref{eq:Hn}). This model approximately  represents interacting cold atoms investigated in  Ref. \cite{Monroe2021observationStarkMBL}.  Another system of $25$ atoms was also considered there. However, the latter system is too complicated for our consideration because of the very large Hilbert space (over $10^{5}$ states with the total spin $1/2$ and fixed dipole moment $P=0$ modulo $25$). The consideration was still possible for the system of  $29$ spins examined earlier, Sec. \ref{sec:ExpQub}, because its effective Hamiltonian possesses a smaller Krylov subspaces due to a short-range interaction there.

\begin{figure}
\includegraphics[scale=0.45]{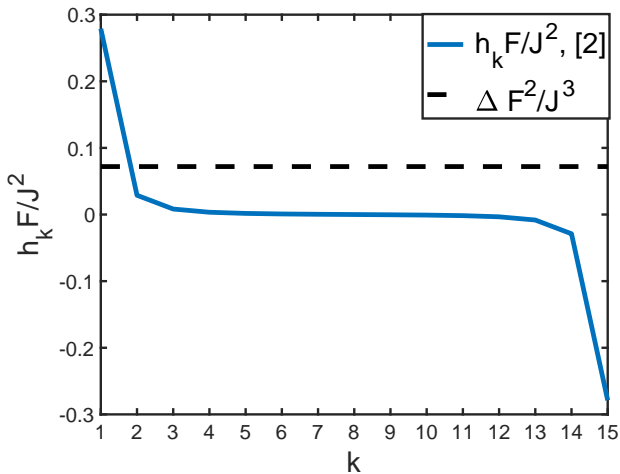} 
\caption{Longitudinal fields vs. the number of spin $k$  generated by means of the Schrieffer-Wolff transformation in the model of Ref. \cite{Monroe2021observationStarkMBL} (blue line)  compared to the four--spin hopping interaction conserving dipole moment (dashed black line).}
\label{fig:hs15}
\end{figure}

Experimentally investigated systems  are characterized by open boundary conditions (OBC). Consequently, similarly to Sec. \ref{sec:ExpQub}, there exist static longitudinal fields  generated in the second order of perturbation theory  Eq. (\ref{eq:sw2nddiag}),    as shown in Fig. \ref{fig:hs15} where the fields are rescaled by the factor $J^2/F$ and compared with four spin hopping amplitude rescaled by the factor $J^3/F^2$  similarly to  that in Fig. \ref{fig:hs}   in  Sec. \ref{sec:ExpQub}. In the case of $J=F$  depicted in  Fig. \ref{fig:hs15},  only fields acting on the  spins at the edges  exceed the four spin hopping amplitude, while for larger field gradient $F$ other fields will be  also significant. Eventually in the large field gradient limit $F\rightarrow \infty$  many-body localization of all states should take place in accord with the experimental observations \cite{Monroe2021observationStarkMBL} similarly to that for Ref. \cite{Guo2020MBLTiltedModel} as it was shown in Sec. \ref{sec:ExpQub}. A  different behavior is expected for the periodic model lacking  longitudinal fields. 



To examine the effect of boundary conditions and the relevance of spin state groups  we  evaluated imbalances exactly for the OBC  Eq. (\ref{eq:Hn})  and  PBC Eq. (\ref{eq:PerH}) models. We use hopping amplitudes $J_{ij}=J/r_{ij}^{1.3}$ as in Ref. \cite{Monroe2021observationStarkMBL} with $r_{ij}= |i-j|$ in the OBC model and $r_{ij}= 2N|\sin(\pi (i-j)/N )|/\pi$ in the  PBC model to make it periodic on a circle. The dependence of group averaged imbalances on the field gradient is investigated for $N=12$ to examine the relevance and applicability of the groups. We also evaluated imbalances for the maximum field gradient $F=2.5J$ used experimentally for the system of $15$ spins and compare the  results with  the periodic system of $14$ spins at the same field  gradient $F=2.5J$. 

The division of states into groups is not formally applicable to the power law  hopping  
$J_{ij}$ since in the limit of a large field gradient four spin hopping Eq. (\ref{eq:4spinint})  can involve arbitrary quartets of spins with transitions conserving dipole moments. However, it can be valid approximately because the dominating hopping is still local due to the fast decrease of its amplitude with the interspin distance. For example the hopping amplitude for the quartet transition $S_{i}^{+}S_{i+1}^{-}S_{i+3}^{-}S_{i+4}^{+}$ of two spin pairs separated by one interatomic distance is less then that for  the local quartet $S_{i}^{+}S_{i+1}^{-}S_{i+2}^{-}S_{i+3}^{+}$ by  almost a factor of $10$ ($8.0$).  

It is not clear whether the power-law hopping within the parent model would lead to the  inevitable localization breakdown in the infinite number of spins limit. According to the preliminary analysis, the hopping $1/r^{1.3}$ in the case of the strong field gradient does not inevitably lead to all state delocalization as in the models of Refs. \cite{ab98book,ab98prl,ab06preprint,Lukin14MBLGen,ab15MBL,ab15MBLXY}. Localization can be unstable for interaction decreasing slower with the distance. Particularly, the recent work \cite{WangStarkMBLInfRange} reports the number of interesting behaviors in the Stark MBL problem in the case of infinite range  distant independent transverse interactions  that awaits the proper interpretation and experimental verification.  It is not clear whether the ergodic spot arguments \cite{Huveneers17BrekDwnLoc} are applicable to the present problem as well because of the lack of quenched disorder.

 \begin{figure}
\includegraphics[scale=0.45]{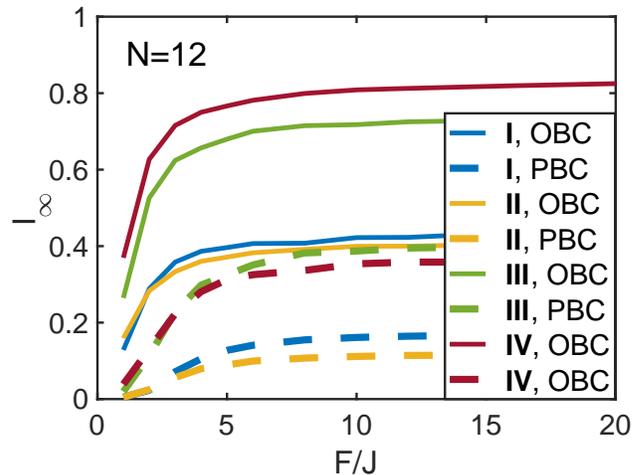} 
\caption{Group averaged  imbalance for the periodic and open boundary condition realizations  of the system of Ref. \cite{Monroe2021observationStarkMBL} vs. the field gradient for $N=12$.}
\label{fig:ImbExN12}
\end{figure}

To examine the applicability of the group concept to the system of interest  we evaluated imbalances for different initial states and average them over initial state groups for $N=12$. The infinite time limit of group averaged imbalances is shown in  Fig. \ref{fig:ImbExN12} for the system with open boundary  conditions   (OBC)  corresponding to the experiment and the periodic system for $12$ spins. In both cases imbalances are distinguishable for different groups already for $F>2J$. This domain includes the experimentally probed field gradients $F\leq 2.5 J$.  As noticed in Sec. \ref{sec:ModelPBC} the dependence of imbalance on the field gradient saturates at $F\geq 4J$. For the OBC problem imbalance slowly increases with the field gradient due to the localization effect of the longitudinal field induced by the boundaries (see Fig. \ref{fig:hs15}). 
 

The behavior of average imbalances is qualitatively similar to that for the  previously considered $29$ spins with a short-range interactions (see Fig. \ref{fig:imbinfq}). Initial states  of the group {\bf IV} correspond to larger imbalances compared to the states of the group {\bf III} in accord with our expectations. For OBC systems the average imbalance increases with increasing the field gradient $F$ towards its maximum value of $1$. 
The minimum imbalance also increases with $F$ indicating  localization of all system states. 


For the experimentally relevant OBC model of $15$ spins and $F=2.5J$  the minimum imbalance $I_{\rm min}=0.1364$ is found for the initial state $\downarrow \downarrow\uparrow\uparrow\downarrow\uparrow\downarrow\downarrow\downarrow\uparrow \uparrow \downarrow \uparrow  \downarrow    \uparrow$. The average imbalances for the groups {\bf III} and {\bf IV} are given by  $0.5221$ and $0.6787$, respectively. This observation suggests the localization of the majority of states in a sharp contrast with the PBC problem for $N=14$ and $F=2.5J$. Group average imbalances for this problem are given by $0.055$, $0.04$, $0.088$, $0.1877$ for the groups {\bf I} - {\bf IV}, respectively.  Thus the minimum imbalance for the OBC problem is  comparable to the averaged imbalance for the most localized group {\bf IV} for the PBC problem. 

This observation suggests delocalization of  states  for the first three groups and localization of most of states belonging to the fourth one. It is consistent with that for the minimalist model  Eq. (\ref{eq:MinimHamCons}) except for the group {\bf III}. However, since there exists non-local  hopping this observation  is approximately  consistent with that in Appendix II (Fig. \ref{fig:NoConstr}). It was shown there that  the addition of the fourth order non-local interaction suppresses localization of  states of the group {\bf III}  for a  field gradient $F<25J$ and even numbers of spins. Consequently, the system containing  even numbers of interacting spins with periodic boundary conditions is most suitable to investigate coexistence of localization and delocalization. 

\section{Discussions and Conclusions} 
\label{sec:Concl}

Here we summarize the results of the present work and compare them with other work \cite{Rakovszky2020,Moudgalya20QHall,Moudgalya2019thermalization,
Bardarson21FragmentedSpaceSampleSt} where the composite blocks (pseudospins, dipoles, defects) were introduced to characterize dipole moment conserving spin dynamics. The results can be divided  into three parts discussed separately including the classification of Krylov subspaces for the minimalist model (Sec. \ref{sec:ConclKr}), understanding spin dynamics in different groups of Krylov subspaces  (Sec. \ref{sec:ConclLoc}) and the application of  these findings to existing experiments  (Sec. \ref{sec:ConclExp}).

\subsection{Understanding Krylov subspaces}
\label{sec:ConclKr}

The visually transparent consideration of spin dynamics in the minimalist periodic dipole  moment conserving (modulo $N$) model for the chain of $N$ spins $1/2$ Eq. (\ref{eq:MinimHamCons}) has been proposed using inverted spin chain with all even numbered  spins overturned about $x$ axis by the angle $\pi$. Then the spin transport for even numbers of spins can be represented as hops of pairs of identically oriented spins by two chain periods to the right or to the left.  This representation is used to separate all product states and associated Krylov subspaces into four groups based on the parity of identically oriented  spin sequences.

\subsubsection{Groups of states} 

The inverted chain states belonging to the group {\bf I} are composed by the only even sequences (see Fig. \ref{fig:Nums}). Corresponding Krylov subspaces are translationally invariant. Consequently,   eigenstates are delocalized. Their dynamics is characterized by the Bethe anzatz integrable anisotropic Heisenberg model for the pseudospins $1/2$ created from adjacent spin pairs similarly to  Refs. \cite{Bergholtz05QHallXYmodel,Moudgalya20QHall}. 


The inverted states belonging to the group {\bf II} have even number of spins in all sequences of adjacent spins oriented either upwards or downwards, while the oppositely oriented sequences must have at least one sequence with more than one spin. Corresponding Krylov subspaces are translationally invariant. Eigenstates belonging to this group are delocalized and we expect them to be ergodic based on the preliminary analysis of the level statistics. 

Several  Krylov subspaces composed by specific combinations of  pseudospins and dipoles belonging to the group {\bf II}  and possessing delocalized eigenstates were identified  in Ref. \cite{Moudgalya2019thermalization}.  The transport visualization using inverted lattice  permitted us to move further and identify all translationally invariant subspaces thus generalizing the previous work. 

The inverted states possessing both upwards and downwards oriented sequences with   odd numbers of spins and systems with odd numbers of spins are characterized by non-translationally invariant Krylov subspaces with marginal exceptions of subspaces containing translationally invariant product states, considered in Ref. \cite{Moudgalya20QHall}. These states can be separated into two groups including the group {\bf III} of states with all mobimle spins and group {\bf IV}, of states possessing ifrozen  spins. All spin sets containing frozen spins are identified.

\subsubsection{Numbers of states} 

The numbers of states belonging to all groups increase exponentially with the numbers of spins $N$. We evaluated this dependence analytically using the generating function method. The number of states increases with the number of the group. The majority of states belong to the group {\bf IV}.


\subsubsection{Integrals of motion}

It is possible to identify a number of conserving quantities similarly to Ref. \cite{Rakovszky2020} where such quantities were considered for the dipole moment conserving transport in the spin $S=1$ chain. This include for instance the number of spin sequences within the representative product state containing an odd number of spins or the parity of the numbers of spins at odd sequence boundaries (the  leftmost spin is always odd, while the rightmost spin  is always even or vice versa). In the present work we do not attempt to identify all such integrals of motion concentrating more on the localization problem. Yet below in Sec. \ref{sec:ConclLoc} we recognize their significance for understanding spin dynamics. 


\subsubsection{Possible extension of the arguments to the spin $S=1$ case.} 

Visualization of spin dynamics within the inverted lattice permits us to represent it  as moving  vehicles, composed by pairs of adjacent, identically oriented spins,  within the environment of the oppositely oriented spins. Similarly, for the minimalist dipole moment conserving spin $1$ problem \cite{Pollmann20Fragment} one can introduce such vehicles made of adjacent spins with opposite projections $\pm 1$. Then these vehicles can move freely  within the environment of the spins with the zero projection. 

It can be shown (we leave the prove to the readers) that with marginal exceptions the product states  forming translationally invariant Krylov subspaces must have spins with the projections $S^{z}= \pm 1$ arranged in the way that  each spin with a projection $S^{z}= \pm 1$   has at least one of the two neighboring spins with  $|S^{z}|= 1$ having the opposite projection $S^{z}= \mp 1$ (that spin can be separated from the given spin  by an arbitrary number of spins with the zero projection).  Further extensions of theory to more complicated models can be possibly developed,  which is beyond the scope of the present work.

\subsection{Localization and delocalization of states in different groups. Connection to the parent model.}
\label{sec:ConclLoc}

It is quite natural to expect delocalization of eigenstates belonging to the translationally invariant groups {\bf I} and {\bf II} and localization for the states of the group {\bf IV} possessing the immobile spins. These expectations are fully confirmed by the investigation of the infinite time imbalances. The study of imbalances also shows the localization of almost all states belonging to the group {\bf III} lacking translational invariance but with all mobile spins. 

The model considered in Refs. \cite{Moudgalya20QHall,Moudgalya2019thermalization,
Bardarson21FragmentedSpaceSampleSt} is different from Eq. (\ref{eq:MinimHamCons}) because it does not include the longitudinal part containing $S^{z}$ operators. Remember that this part of the Hamiltonian is the outcome of generalized Schrieffer Wolff expansion of the parent XY model, which makes our model more relevant experimentally. To the best of our knowledge in other systems the strong longitudinal interaction is always unavoidable. 

To compare system dynamics in  two different models we evaluated average imbalances for all groups of states for the model of Refs.  \cite{Moudgalya20QHall,Moudgalya2019thermalization,
Bardarson21FragmentedSpaceSampleSt}. It turns out that the imbalances behave nearly identically for all groups except for the group {\bf III}, where the ergodic behavior is found in contrast to the localization in the model Eq. (\ref{eq:MinimHamCons}), considered in the present work.  

What is the origin of this difference? In our opinion the longitudinal interaction acts as a quasi-static  disorder, localizing spin dynamics. The static nature of disorder can be the consequence of the conserving quantities (cf. Ref. \cite{Rakovszky2020}) including  the number of odd sequences and positions of their borders that can occupy limited number of places due to the translational invariance breakdown. It is not clear whether this localization survives for group {\bf III} states in a thermodynamic limit of an infinite system. At least no signature of localization breakdown is seen in Fig. \ref{fig:Imb} up to $N=30$.

The localizing effect of the longitudinal interactions should be even stronger in other systems  including   the fractional quantum Hall problem in the thin-torus limit  \cite{Bergholtz05QHallXYmodel,SeidelQHall05,Wang12QHallDipCons,Moudgalya20QHall}  and 
the anisotropic Heisenberg model with nearest neighbor interactions subjected to a uniformly growing field  in a large  field gradient limit \cite{Bardarson21FragmentedSpaceSampleSt,Pollman2021StarkMBLOpen,Taylor20ExpShatt}. In both limits justifying the transverse interaction in the form of Eq. (\ref{eq:MinimHamCons}) there exist longitudinal interactions exceeding the transverse interaction by the expansion parameter. This should increase the strength of static disorder and the localization trend. Consequently, the present model is expected to be most delocalized, that justifies its experimental and theoretical considerations.  

The concept of the groups of states is extendable at least qualitatively to the parent models with time periodic Hamiltonians that can be used to realize the minimalist model experimentally for the large field gradient $F$. It is possible  that localization survives there for $F>J$ because of the quasistatic disorder induced by some integrals of motion including for example the approximately conserving dipole moment. This expectation is consistent with the recent observation  of Stark gauge protection \cite{Haifeng2022GaugeInvStarkMBL}, where the disorder free localization in several quantum gauge theory realizations   has been stabilized by the uniformly growing potential  in spite of  the presence of gauge-breaking interactions.

\subsection{Interpretation of experiments and suggestion of their advancement}
\label{sec:ConclExp}

The localization observed experimentally in Refs. \cite{Guo2020MBLTiltedModel,Monroe2021observationStarkMBL} for the large field gradient $F \sim 2.5 J$ turns out to be the consequence of the boundaries and lack of certain interactions in Refs. \cite{Guo2020MBLTiltedModel}. This issue can be resolved using the periodic spin chain under action of a time–varying Hamiltonian suggested within the present work. Then the system becomes translationally invariant and it should show a full diversity of behaviors depending on the group of the initial state. 

The initial product states possessing the minimum imbalances are determined for various experimental settings. It can be worth to probe these substantially delocalized regimes experimentally. 



\begin{acknowledgments} 
{\it Acknowledgement.} A. B. acknowledges the support by Carrol Lavin Bernick Foundation Research Grant (2020-2021), NSF CHE-2201027
grant and LINK Program of the NSF and Louisiana Board of Regents.
\end{acknowledgments}.

\bibliography{MBL}

\appendix

\begin{widetext}

\section{Applicability of the main approximation}
\label{sec:AppRelev}

Here we  in greater detail the relevance of the minimalist  model for characterizing system dynamic. The minimalist model periodic in time and space is derived in the main text using the generalized 
Schrieffer-Wolff  transformation of the time periodic $XY$ model subjected to the uniformly growing field as the first non-vanishing third order term of the effective Hamiltonian expansion in the ratio of interspin interaction $J$ and field gradient $F$ (see Sec. II.B of the main text). This expansion should be valid at sufficiently large field gradient $F > F_{c}$, where the critical field gradient can depend on the number of spins $N$. 

In Sec. II.C the conservative and liberal approaches to the estimate of $F_{c}$ are suggested. The conservative estimate requiring the accurate estimate of eigenstate energies within the minimalist model  results in exponentially large (in $N$)  field $F_{c1} \sim J2^{N/2}$. The liberal approach requiring  the convergence of the Schrieffer-Wolff  expansion suggests $F_{c2} \sim J$.  

Below we analyze the convergence of the generalized Schrieffer-Wolff  transformation deriving the fourth order correction to the effective Hamiltonian in Sec. \ref{sec:modPBC4th} and verifying the convergence of the generalized Schrieffer-Wolff  transformation in Sec. \ref{sec:SchrWConv}. Then eigenstate energies obtained using the minimalist model and corrected Hamiltonian are compared to each other in Sec. \ref{sec:Relev}. This comparison results in the estimate of the critical field gradient $F_{c}$  approximately consistent with the conservative estimate. In addition we compare average imbalances for different groups of initial states evaluated using the minimalist model and the original time periodic problem.  This study shows that the imbalance behavior for the minimalist model is qualitatively consistent with the exact system already for $F>F_{c2}$ though the quantitative agreement of two results for localized states of the third and fourth groups requires $F>F_{c1}$.

All comparisons are performed for the $XY$ model with nearest and next neighbor interactions different from zero and identical to each other similarly to the experimentally investigated system in Ref. \cite{Guo2020MBLTiltedModel}.  
 We set both interactions equal to $J$.


\subsubsection{The fourth order correction to the effective Hamiltonian.}
\label{sec:modPBC4th}

The fourth order correction is considered to validate the relevance of the minimalist model and the generalized Schrieffer-Wolff  transformation. It consists of two contributions including the fourth order contribution from the    Baker-Campbell-Hausdorff-Dynkin formula $\widehat{H}_{4a}$ and the higher order in $F^{-1}$ Schrieffer-Wolff  transformation contribution  $\widehat{H}_{4b}$ originated from  the second order transverse interaction, not conserving dipole moment, emerging similarly to the one for the OBC problem (see Eq. (\ref{eq:sw2ndoffdiag}) in Sec. \ref{sec:ModelOBC}  in the main text) that can be written as   
\begin{eqnarray}
\widehat{H}_{2offd}
=-\sum_{j, k~= l}2A_{jk}A_{lj}e^{2\pi i (k-j)/N}S_{j}^{z}S_{l}^{+}S_{k}^{-}+\sum_{j~=m, k}2A_{jk}A_{km}e^{2\pi i(k-j)/N}S_{k}^{z}S_{j}^{+}S_{m}^{-}.
\label{eq:H2OffRes}
\end{eqnarray}
Its contribution to the effective Hamiltonian  can be derived introducing operators $\widehat{S}_{2}$ and  $\widehat{S}_{2mod}$ similarly to Sec. \ref{sec:ModelPBC} (see Eq. (\ref{eq:4thOrdTerm}) below). 

Then the fourth order correction to the effective Hamiltonian can be expressed as (cf. Eq. (\ref{eq:BCHExp}))
\begin{eqnarray}
\widehat{V}_{4}=\left[\widehat{H}_{4a}+\widehat{H}_{4b}\right]_{P},
\nonumber\\
\widehat{H}_{4a}=\frac{i}{8\tau_{F}}[\widehat{S}_{mod},[\widehat{S}_{mod},[\widehat{S}_{mod},\widehat{S}_{1}]]]+\frac{i}{12\tau_{F}}[\widehat{S}_{1},[\widehat{S}_{mod},[\widehat{S}_{mod},\widehat{S}_{1}]]]
\nonumber\\
+\frac{i}{24\tau_{F}}[\widehat{S}_{mod},[\widehat{S}_{1},[\widehat{S}_{mod},\widehat{S}_{1}]]]+\frac{i}{24\tau_{F}}[\widehat{S}_{1},[\widehat{S}_{1},[\widehat{S}_{mod},\widehat{S}_{1}]]],
\nonumber\\
\widehat{H}_{4b}=\frac{1}{2}[\widehat{S}_{2mod}, \widehat{H}_{2offd}],
\nonumber\\
\widehat{S}_{2mod}=\frac{1}{16F}\sum_{j\neq l, k}\frac{2\pi i (j+l-2k)J_{jk}J_{kl}S_{k}^{z}S_{j}^{+}S_{l}^{-}}{2NF(j-k)(l-k)}\frac{e^{2\pi i\frac{j-l}{N}}}{1-e^{2\pi i\frac{j-l}{N}}}.
\label{eq:4thOrdTerm}
\end{eqnarray}
Remember that  the subscript $P$ means that the only terms conserving dipole moment $P$ (modulo $N$) are  left. 

\subsubsection{Convergence  of the perturbation series.}
\label{sec:SchrWConv}

Here we investigate the convergence of the expansion of the effective Hamiltonian defined by  Eq. (\ref{eq:PulseTarget}) in the main text. To validate the expansion convergence  we compare the third and fourth order approximations to the effective Hamiltonian with the results of the exact diagonalization of the evolution matrix Eq. (\ref{eq:pulsetot}) that can be performed at relatively small numbers of spins $N\leq 15$. 

 The target Hamiltonian $\widehat{H}_{eff}$ can be considered in the third order approximation 
referred as the {\it minimalist} model and in the fourth order approximation with the fourth order term  Eq. (\ref{eq:4thOrdTerm}) added to the minimalist model Hamiltonian., The fourth order approximation   is referred as the {\it improved} model. The  minimalist and improved models  are compared with the results of the exact diagonalization of the evolution matrix. Remember that here and in the next section all calculations are performed for the parent $XY$ model with nearest and next-neighbor interactions $J_{1}$ and $J_{2}$. We also set $J_{1}=J_{2}=J$ similarly to the interactions in Ref. \cite{Guo2020MBLTiltedModel}.  

To compare the approximate and exact models we performed exact diagonalization of the minimalist and corrected model Hamiltonians, and exact diagonalization   of the evolution matrix (see Eq. (\ref{eq:pulsetot}) in the main text). Its eigenvalues are complex numbers $z_{i}$ with absolute value unity as it has to be for the unitary matrix. According to the definition off the effective Hamiltonian (see Eq. (\ref{eq:PulseTarget}) in the main text)  logarithms of those numbers for the specific eigenstate $a$ possessing the dipole moment $P_{a}$ can be expressed as 
\begin{eqnarray}
\ln(z_{i})=-i\tau_{F}(-FP_{a}+E_{a*}), 
\label{eq:EigU}
\end{eqnarray}
where $E_{a*}$ is the eigenenergy of the corresponding state $a$ of the effective  Hamiltonian $\widehat{H}_{eff}$. This definition makes sense only for sufficiently large field gradient $F\gg J$. The dipole moment is approximately conserved under this condition.  It turns out that already  for $F>2.5J$  ($N\leq 14$) imaginary parts of eigenvalue logarithms form $N$ well separated groups corresponding to certain dipole moments (mod $N$). Consequently,  target energies $E_{a*}$ are clearly identifiable for each dipole moment.

\begin{figure}
\includegraphics[scale=0.5]{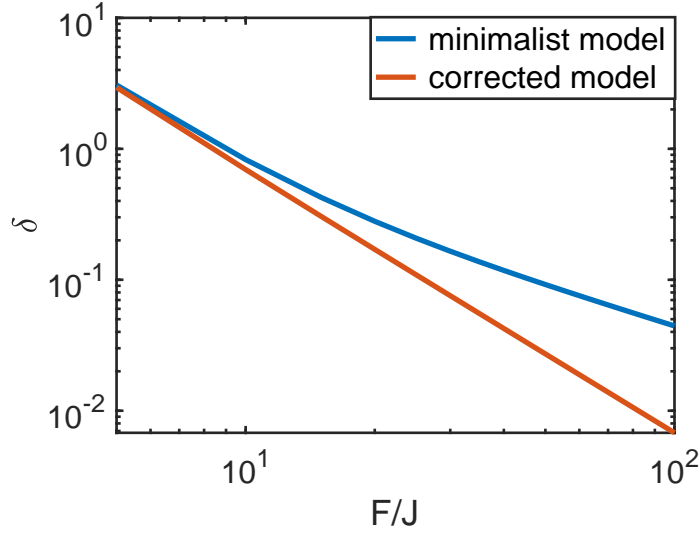} 
\caption{Relative deviation Eq. (\ref{eq:DevDef})  of eigenstate energies  evaluated within the minimalist and corrected models from exact eigenenergies.}
\label{fig:ErrEx}
\end{figure}

Then we compare these energies with eigenenergies of the effective Hamiltonians of the minimalist and improved models. To compare eigenenergies   we evaluated the relative deviation $\delta$ of two ordered sets of eigenenergies, which is defined as 
\begin{eqnarray}
\delta =\frac{1}{Z} \sum_{i=1}^{Z}\frac{(E_{i,1}-E_{i,2})^2}{\delta E^2}, 
\label{eq:DevDef}
\end{eqnarray} 
where $Z$ is the total number of eigenstates, $E_{i,1}$, $E_{i,2}$ are  eigenenergies of the state $i$ obtained using methods $1$ and $2$  and $\delta E$ is the average energy splitting between adjacent energy levels, which was  calculated using the most accurate approximation.

Calculations of relative deviations at different field gradients $F$ show that the deviations of the  minimalist and improved models from the exact results scale with $F$ as $1/F$ and $1/F^{2}$, respectively,  as illustrated in Fig. \ref{fig:ErrEx} for the states of $N=12$ spins with a zero  total spin projection to the $z$ axis and zero dipole moment, $P=0$ (mod $N$). Calculations for other states show a similar behavior.  Consequently, the suggested expansion of the effective Hamiltonian converges at large field gradients $F$, which proves its relevance.  

\subsubsection{Relevance of the main approximation.}
\label{sec:Relev}

The Hamiltonian of the minimalist model considered in the main text  represents the  first non-vanishing term in the  effective Hamiltonian expansion  in $JF^{-1}$  Eq.  (\ref{eq:Heff3rd}).   This approximation  should be valid at sufficiently large field gradient $F>F_{c}$. Here we estimate the minimum gradient $F_{c}$ where this approximation is still applicable by comparing energy levels Eq. (\ref{eq:DevDef})  and imbalances (Eq. (\ref{eq:ImbDefMT})  in the main text)  calculated within this approximation (minimalist model of Eq. (\ref{eq:MinimHamCons})) and with the fourth order correction of Eq. (\ref{eq:4thOrdTerm}) added ({\it improved} model).

Eigenenergies obtained using two different approaches  are compared using  Eq. (\ref{eq:DevDef}) for even numbers of spins $10\leq N \leq 18$ and field gradients $5<F<100$. The results for states with zero spin projection to the $z$ axis and minimum absolute value of the dipole moment ($P=0$ for $N=12$, $16$ and $P=0.5$ for $N=10$, $14$, $18$) are reported in Fig. \ref{fig:ErrAppr}. 

\begin{figure}
\includegraphics[scale=0.5]{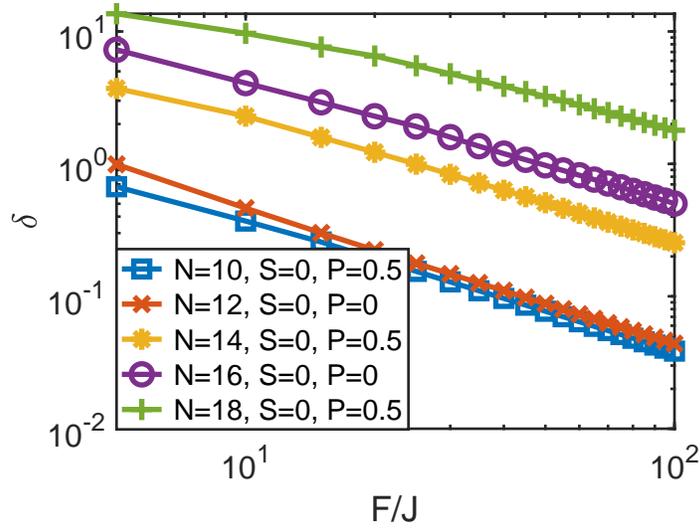} 
\caption{Relative deviation of eigenenergies Eq. (\ref{eq:DevDef}), evaluated within the minimalist and corrected models, from each other.}
\label{fig:ErrAppr}
\end{figure}

We set the formal criterion for the validity of the minimalist model to be $\delta=1$ in Eq. (\ref{eq:DevDef}).  According to this criterion  one can estimate  $F_{c}/J \approx 5$ for $N=10$ and $12$, $F_{c}/J \approx 30$ for $N=14$, $F_{c}/J \approx 60$ for $N=16$ and $F_{c}/J > 100$ for $N=18$. The observed trend for largest numbers of spins is approximately consistent with the conservative estimate of the crossover field gradient  obtained in the main text. One can approximately describe that crossover as 
\begin{eqnarray}
F_{c1}(N) \approx 0.25\cdot 2^{N/2}J,
\label{eq:FcEst}
\end{eqnarray} 
where the factor $0.25$ is chosen  to satisfy the condition $\delta \approx 1$ in Fig. \ref{fig:ErrAppr} for $N=16$. 
 
 The minimalist model Eq. (\ref{eq:MinimHamCons}) can have a wider applicability domain since it can leave  average properties (e. g. imbalances) unchangeable in spite of large shifts of energy levels. To partially address this concern  we consider the difference of imbalances evaluated in the Minimalist  Model  and  in the original time periodic model (Eq. (\ref{eq:PerH}) in the main text). Both imbalances are evaluated within the infinite time limit. In both cases we use the most straightforward definition of imbalance (see Eq. (\ref{eq:ImbDefMT}) in the main text) defining the spin projection expectation values to be equal to average spin projection to the $z$ axis  since our goal here is to examine the  relevance of the perturbation theory. 
  




\begin{figure}[h]
\centering
\subfloat[]{\includegraphics[scale=0.4]{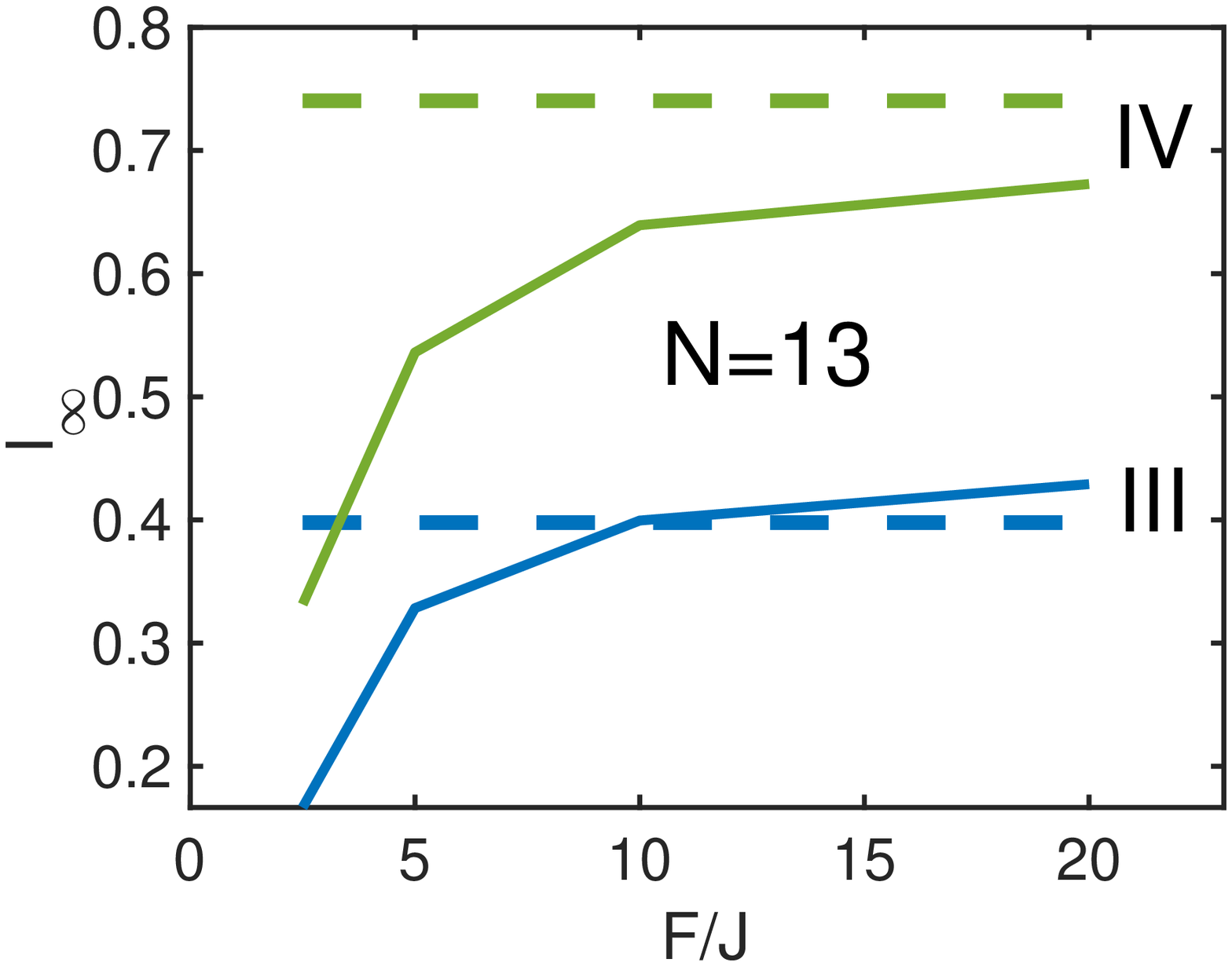}} 
\subfloat[]{\includegraphics[scale=0.4]{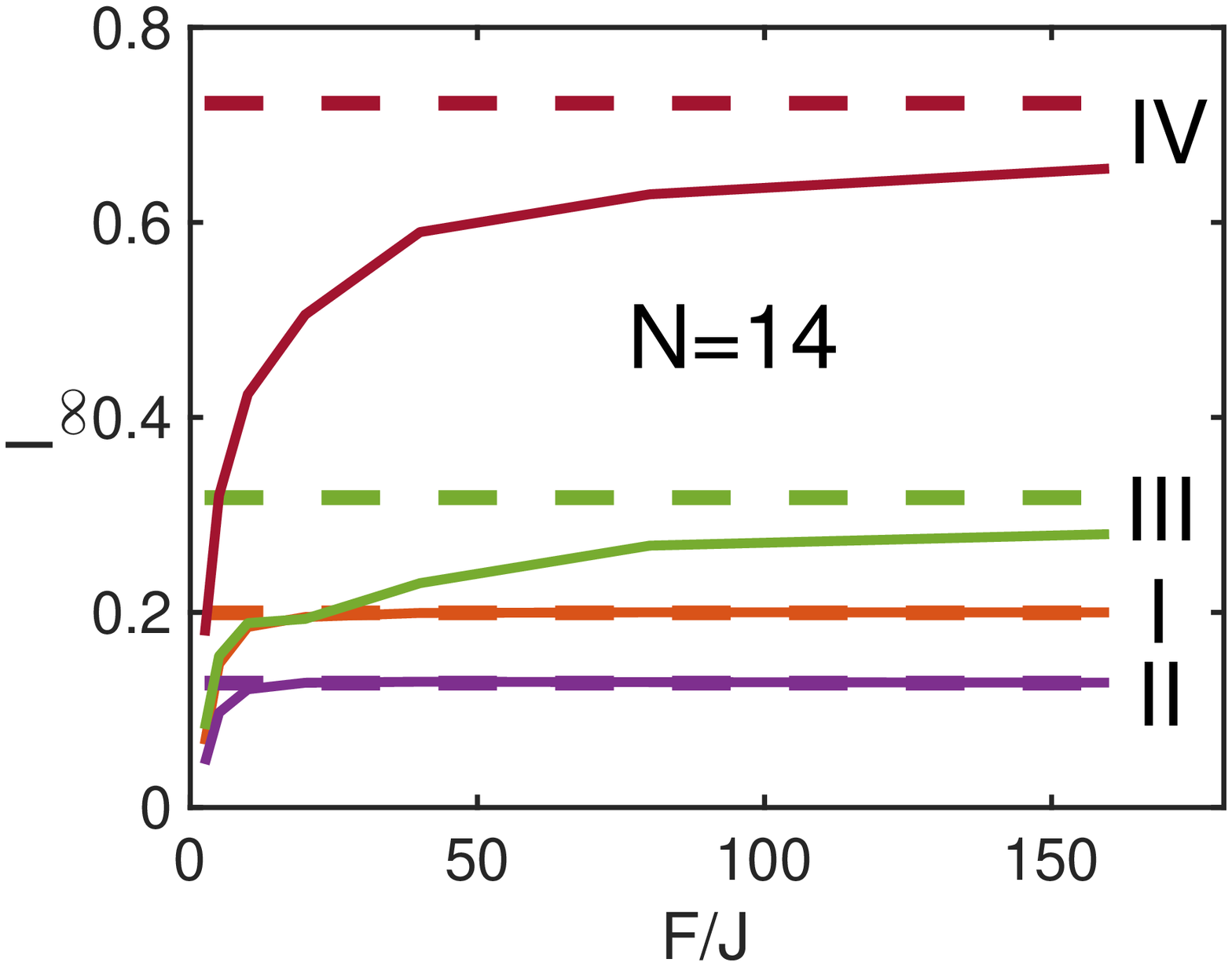}}
\caption{\small  Group averaged imbalance dependence on the field gradient for $N=13$   ({\bf a})  and  $N=14$ ({\bf b}).  Group is indicated by the Roman number at the right side of the graph. Dashed lines show imbalances evaluated within the minimalist model. 
} 
\label{fig:NoConstr}
\end{figure}

 In Fig. \ref{fig:NoConstr} the  comparison of imbalances evaluated within two models is reported for $N=13$ and $N=14$. Dashed lines show the results for the MM model  Eq. (\ref{eq:MinimHamCons}), which are independent of the field gradient, while solid lines show the results for the original time-periodic model Eq. (\ref{eq:PerH}). 
 
Obviously, the minimalist model is not applicable at $F \sim J$ where imbalances are approximately group independent.  As it is shown in Fig. \ref{fig:NoConstr}  the remarkable difference between imbalances evaluated for different groups  is already seen at $F=2.5J$. The field gradient dependence of imbalances for groups {\bf I} and {\bf II} for $N=14$  and all groups for $N=13$ saturates at around $F \sim 5J$. This is consistent with the liberal criterion $F_{c} \sim J$. 
 

The average imbalances for groups  {\bf III} and {\bf  IV}  and $N=14$  reach the  minimalist model  limit at larger fields $F \sim 40 J$ that is consistent with the conservative  estimate Eq. (\ref{eq:FcEst}). Yet for the group {\bf IV} imbalance exceeds other group imbalances and shows the localization trend already for $F \geq 5J$,  while the states of group {\bf III} can be delocalized for $F<40J$. The similar difference between even and odd size behaviors is seen for larger numbers of spins $N=15$ and $16$ as well. 

\section{Calculation of numbers of states in different groups.}
\label{sec:GenerFunc}


Here we report the analytical calculations of the numbers of states belonging to  the groups {\bf II}  (Sec. \ref{sec:GenerFuncGrII})   and  {\bf III} (Sec. \ref{sec:GenerFuncGrIII}) using generating function method, while the number of states belonging to the group {\bf I} ($2\cdot 2^{N/2}$) is given  in the main text and the number of states belonging to the group {\bf IV} approaches the total number of states $2^{N}$. These results are summarized in Fig. \ref{fig:Nums} in the main text. Also the numbers of states with all frozen spins as a function of the total number of spins  is evaluated  in Sec. \ref{sec:GenerFuncImm} (this number is also quoted in the main text). 

\subsection{Counting group {\bf II} states.} 
\label{sec:GenerFuncGrII}

According to the main text the states of group {\bf II} can be represented by states having even numbers of sequences with either all odd or all even-numbered sequences containing even numbers of spins. Each state must have at least one odd sequence to be distinguished from group {\bf I} states having all even sequences. Also completely frozen states having either all odd or all even-numbered  sequences containing  only one spin should be excluded. 

States of periodic system  belonging to the group {\bf II} can be distinguished by states of the first and last ($N^{th}$) spins as shown below
\begin{eqnarray}
({\rm \mathbf{A}})~ \uparrow\uparrow\downarrow\uparrow\uparrow\uparrow\uparrow\downarrow\downarrow\downarrow, ~~~ ({\rm \mathbf{B}})~ \uparrow\downarrow\uparrow\uparrow\uparrow\uparrow\downarrow\downarrow\downarrow\uparrow. 
\label{eq:GroupII2opt}
\end{eqnarray} 
 The states of the type $A$ has first and last spins different, while they are identical for the states of the type $B$ representing the circular shifted state $A$. It is more convenient to calculate first the number of states of the type $A$ and then calculate the number of states of the type $B$ representing them in terms of the states of the type $A$ of the smaller size  ($8$ spins in the three internal sequences  within  the state $B$ in Eq. (\ref{eq:GroupII2opt})) with the remaining sequence being added to both sides.

We begin with the calculation of the number of states of the type $A$ having even number of sequences (because the first and last spins are opposite to each other) and having even numbers of spins  either in all odd or even-numbered sequences. This definition  includes all states of the group {\bf I} containing only even sequences and fully frozen states belonging to the group {\bf IV} having a single spin in all sequences in odd or even-numbered positions. They will  be excluded from the final answer. 

 Let $W_{N}^{oo}$ and $W_{N}^{eo}$ be the numbers of states of $N$ spins  containing  odd or even numbers of sequences, respectively, and having all odd-numbered  sequences containing even numbers of spins  (e. g. states with $N=9$ represented as $(2,5,2)$ or $(2,3,2,2)$ treated as with the open boundary conditions, i. e.  distinguishing the first and the last sequences). Similarly, $W_{N}^{oe}$ and  $W_{N}^{ee}$ are the numbers of states of $N$ spins containing  odd or even numbers of sequences, respectively, having all even-numbered sequences containing even numbers of spins.  Then the state of  $N > 0$ spins  can be obtained from the smaller state by means of adding the sequence  of identical spins to the right with their projections  opposite to those in the last sequence.  We assume no sequences with negative numbers of spins and one sequence with $0$ spins. 
 
 The states with $N$ spins, having odd numbers of sequences and all odd-numbered sequences possessing even numbers of spins can be generated from the states having even numbers of sequences and even numbers of spins in odd-numbered sequences by means of adding one more sequence (to the right), having even numbers of spins $2k$ ($k=1, 2, 3,...$). If $N$ is even this state can be also obtained as one of two states of all spins oriented upwards or downwards added to the zero spin state. Formally this condition can be written as 
\begin{eqnarray}
W_{N}^{oo}=\sum_{k=1}^{\infty}W_{N-2k}^{eo}(1+\delta_{N-2k,0}),
\label{eq:evoo}
\end{eqnarray}
where $\delta_{ab}$ is the Kronecker symbol.  

The states of $N$ spins, having even numbers of sequences and all odd-numbered sequences possessing even numbers of spins can be generated from the states having odd number of sequences and even number of spins in odd-numbered sequences by means of adding one more sequence, having any numbers of spins $k>0$. Consequently, we get 
\begin{eqnarray}
W_{N}^{eo}=\sum_{k=1}^{\infty}W_{N-k}^{oo}+\delta_{N0}. 
\label{eq:eveo}
\end{eqnarray}
The last term $\delta_{N0}$ in the right hand side of Eq. (\ref{eq:eveo}) reflects the fact that there is one state with even number zero of sequences for $N=0$. 
 
Using similar arguments one can obtain equations for state numbers $W_{N}^{oe}$, $W_{N}^{ee}$ having odd or even numbers of sequences with even sequences in even-numbered positions in the form
\begin{eqnarray}
W_{N}^{oe}=\sum_{k=1}^{\infty}W_{N-k}^{ee}(1+\delta_{N-k,0}), ~ 
W_{N}^{ee}=\sum_{k=1}^{\infty}W_{N-2k}^{oe}+\delta_{N0}. 
\label{eq:eveoe}
\end{eqnarray}
The number of states $W_{A}(N)$ for $N$ spins in sequences of the type $A$ in Eq. (\ref{eq:GroupII2opt}) is the sum of two out of four  above defined numbers for total even number of sequences ,  i. e. $W_{A}(N)=W_{N}^{eo}+W_{N}^{ee}$. 
 
The solution of all equations can be found using generating functions defined as 
\begin{eqnarray}
w_{ab}(x)=\sum_{k=0}^{\infty}W_{N}^{ab}x^{N}. 
\label{eq:GenFunc}
\end{eqnarray}
Eqs. (\ref{eq:evoo}), (\ref{eq:eveo}), (\ref{eq:eveoe}) can be rewritten for these generating functions as 
\begin{eqnarray}
w_{oo}=\frac{x^2}{1-x^2}w_{eo}+\frac{x^2}{1-x^2}, ~ w_{eo}=\frac{x}{1-x}w_{oo}+1, 
\nonumber\\
w_{oe}=\frac{x}{1-x}w_{ee}+\frac{x}{1-x}, ~ w_{ee}=\frac{x^2}{1-x^2}w_{oe}+1. 
\label{eq:GenFuncEq}
\end{eqnarray}
The numbers of interest ($W_{A}(N)$) are defined by the coefficients of the power series expansion 
\begin{eqnarray}
w_{A}(x)=\sum_{N=0}^{\infty}W_{A}(N)x^{N}=w_{eo}(x)+w_{ee}(x)-w_{e}(x),  
\label{eq:powser}
\end{eqnarray}
where $w_{e}(x)$ is the generating function for the number of states of $N$ spins composed by an even number of sequences all having even numbers of spins $w_{e}(x)=\sum_{N}W^{e}_{N}x^{N}$.  It has to be subtracted because these states are included twice both into numbers  $W_{N}^{eo}$ and $W_{N}^{ee}$. 

The functions $W^{e}_{N}$, $W^{o}_{N}$ representing the numbers of states composed by even or odd numbers of sequences containing even numbers of spins can be evaluated also using generating function method. Their generating functions are given by 
\begin{eqnarray}
w_{o}(x)=\frac{2x^2(1-x^2)}{1-2x^2}, ~~ w_{e}(x)=\frac{2x^4}{1-2x^2}. 
\label{eq:Gr1GenF}
\end{eqnarray}
The sum $w_{o}(x)+w_{e}(x)$ yields $1/(1-2x^2)$ which is the generating function for the number of spin pair states $2^{N/2}$.  


Solving Eqs. (\ref{eq:GenFuncEq}) we get 
\begin{eqnarray}
w_{A}(x)=2+\frac{4x^3}{1-x+x^2}-\frac{2x^4}{1-2x^2}; 
\label{eq:powserAns}
\end{eqnarray}

Using the power serious expansion (it is straightforward after the partial fraction expansion of the answer) we got the numbers of  states of the type $A$ for $N>0$ in the form  $(N\geq 2)$
\begin{eqnarray}
W_{A}(N)=8\frac{\sqrt{5}}{5}\left[\frac{\sqrt{5}-2}{\sqrt{5}-1}\left(\frac{\sqrt{5}+1}{2}\right)^{N}-(-1)^{N}\frac{\sqrt{5}+2}{\sqrt{5}+1}\left(\frac{\sqrt{5}-1}{2}\right)^{N}\right]-2^{N/2}/2 +\delta_{N,2}. 
\label{eq:AnsConsFrSt}
\end{eqnarray}
Consequently, in a limit of a large number of spins one can use the exponential asymptotic behavior $W_{A}(N) \propto \lambda_{min}^{-N}$, where $\lambda_{min} = (\sqrt{5}-1)/2 \approx 0.618$ is the root     of the denominators in  Eq. (\ref{eq:powserAns}) possessing the minimum absolute value.   This exponential asymptotic  is valid for the whole number of states in group {\bf II} since it should not depend on boundaries as we will see below. Subtraction of group {\bf I} states will not affect this behavior because their number increases with $N$ slower (as $2^{N/2}\approx 1.41^{N}$). 

To evaluate the number of states of the type $B$ one can represent each of them as the state of the type $A$ of the smaller size $N-z$ and consider all $z-1$ ways of appending the additional sequence ($(2k-1)(W_{N-2k}^{oo}+W_{N-2k}^{oe}-W_{N-2k}^{o})$ for even length $z=2k$ of the additional sequence or $2kW_{N-2k-1}^{oo}$ for its odd length $z=2k+1$) for all possible numbers $z$. The number of states with odd number of sequences  all containing even numbers of spins is subtracted from the definition to avoid including it twice. 

Then one can express the number of states of the type $B$ as 
\begin{eqnarray}
W_{B}(N)=\sum_{k=1}^{\infty}\left[(2k-1)(W_{N-2k}^{oo}+W_{N-2k}^{oe}-W_{N-k}^{o})+2kW_{N-2k-1}^{oo}\right]. 
\label{eq:StB1}
\end{eqnarray}
The associated generating function $w_{B}(x)$ can be expressed as 
\begin{eqnarray}
w_{B}(x)=(w_{oo}(x)+w_{oe}(x)-w_{o}(x))\frac{x^2+x^4}{(1-x^2)^2}+w_{oo}(x)\frac{2x^3}{(1-2x^2)}. 
\label{eq:GenFWB}
\end{eqnarray}

To find the total number of states belonging to the group {\bf II} we need to add both contributions $w_{A}(x)$ and $w_{B}(x)$   found above and subtract all states of the group {\bf I} having the number $2\cdot 2^{N/2}-4$ in accordance with the main text and fully frozen states  $W_N^{II, im}$ evaluated below in Eqs. (\ref{eq:GrIIIm}) and (\ref{eq:Gr2ImFin}). Then the generating function for the total number of states within the group {\bf II} takes the form 
\begin{eqnarray}
w_{II}(x)=w_{A}(x)+w_{B}(x) -\frac{2}{1-2x^2}+\frac{4}{1-x^2}- w_{II,im}=\frac{2(4x^5+12x^4-12x^2+3)}{2x^5+4x^4-x^3-4x^2+1}-\frac{2x^3(2+x)}{1-2x^2}. 
\label{eq:GrIIAns}
\end{eqnarray}
The expansion of the generating function  $w_{II}(x)$ into power series results in the exact answer below for the numbers of states belonging to the group {\bf II}
\begin{eqnarray}
W_{N}^{II}=2\left[\left(\frac{\sqrt{5}+1}{2}\right)^N+(-1)^{N}\left(\frac{\sqrt{5}-1}{2}\right)^N -2^{N/2}\left(\frac{5+2\sqrt{2}}{4} +(-1)^{N}\frac{5-2\sqrt{2}}{4}\right) +\delta_{2,N} +2\right]. 
\label{eq:GrIIFinAns}
\end{eqnarray}
This answer is applicable  only for even numbers of spins $N$.

\subsection{Counting completely frozen states.}
\label{sec:GenerFuncImm}

Completely frozen  states are formed by spin sets  consisting of all identical spins (trivial case of two states) or they must have each sequence containing more than one spin being surrounded by sequences containing only one spin. 
To solve this problem for a periodic system we can use the approach of Eq. (\ref{eq:GroupII2opt}) splitting all states into types $A$ and $B$  based on the first and the last spins. Then the total number of states of interest can be expressed as 
\begin{eqnarray}
W_{N}^{im} = W_{N}^{A} + W_{N}^{B}, ~ W_{N}^{B}=\sum_{k=2}^{\infty}Y_{N-k}^{o1}(k-1), 
\label{eq:DefImmSt}
\end{eqnarray}
where $W_{N}^{A}$ is the number of frozen states of type $A$ with $N$ spins, $W_{N}^{B}$ is the number of $N$ spin frozen states of type $B$ with $N$ spins and $Y_{N}^{o1}$ is the number of frozen states containing an odd number of sequences with first and last sequence containing only one spin.  There is no other way to compose the state of type $B$ avoiding mobile spins in the edge sequence. One can reexpress Eq. (\ref{eq:DefImmSt}) in terms of the corresponding generating functions as (cf. Eq. (\ref{eq:GrIIAns})) 
 \begin{eqnarray}
w_{im} = w_{A} + \frac{x^2}{(1-x)^2}y_{o1}.
\label{eq:DefImmStGF}
\end{eqnarray}

To find numbers of specific states we introduce the functions $Y_{N}^{o1}$,  
$Y_{N}^{o}$, $Y_{N}^{e1}$,  $Y_{N}^{e}$  and $W_{N}^{o1}$,  
$W_{N}^{o}$, $W_{N}^{e1}$,  $W_{N}^{e}$. The states denoted by the letter $Y$ begins with the sequence containing one spin, while the states denoted by the letter $W$ have the initial sequence of more than one spins. Superscripts $o$ or $e$ mean that the states contain odd or even numbers of sequences, respectively. Superscript $1$ means that the considered states are ended by the sequence containing one spin. With these definitions one has $W_{N}^{A}=W_{N}^{e1}+Y_{N}^{e1}+W_{N}^{e}+Y_{N}^{e}$ and, consequently, $w_{A}=w_{e1}+y_{e1}+w_{e}+y_{e}$. Finding all eight functions resolves the problem of interest. 

The functions of interest satisfy the equations describing their evolution with increasing the number of spins $N$ as 
\begin{eqnarray}
Y_{N}^{o}=\sum_{k=2}^{\infty}Y_{N-k}^{e1}, ~ Y_{N}^{e}=\sum_{k=2}^{\infty}Y_{N-k}^{o1}, ~ Y_{N}^{o1}=  Y_{N-1}^{e}+ Y_{N-1}^{e1}+ 2\delta_{N1}, ~ Y_{N}^{e1}=  Y_{N-1}^{o}+ Y_{N-1}^{o1},
\nonumber\\ 
W_{N}^{o}=\sum_{k=2}^{\infty}W_{N-k}^{e1}+2(1-\delta_{N,1}), ~ W_{N}^{e}=\sum_{k=2}^{\infty}W_{N-k}^{o1}, ~ W_{N}^{o1}=  W_{N-1}^{e}+ W_{N-1}^{e1}, ~ W_{N}^{e1}=  W_{N-1}^{o}+ W_{N-1}^{o1}. 
\label{eq:totImEv}
\end{eqnarray}  
These equations can be reformulated in the algebraic form for the generating functions 
\begin{eqnarray}
y_{o}=\frac{x^2}{1-x}y_{e1}, ~ y_{e}=\frac{x^2}{1-x}y_{o1}, ~ y_{o1}=xy_{e}+xy_{e1}+2x, ~ y_{e1}=xy_{o}+xy_{o1}, 
\nonumber\\
w_{o}=\frac{x^2}{1-x}w_{e1} +\frac{2x^2}{1-x}, ~ w_{e}=\frac{x^2}{1-x}w_{o1}, ~
w_{o1}=xw_{e}+xw_{e1}, ~ w_{e1}=xw_{o}+xw_{o1}. 
\label{eq:totImEv1}
\end{eqnarray} 
Solving these equations we obtained the generating function in Eq. (\ref{eq:DefImmStGF}) in the form 
\begin{eqnarray}
w_{im}=\frac{2x^2(x^5 - 2x^4 + x^3 - 2x^2 + 1)}{(1-x)(x^6 + x^4 - 2x + 1)}.
\label{eq:GenFuncImmAns}
\end{eqnarray}
The exponential asymptotic $W_N \propto \lambda_{min}^{-N}$ is determined by the root $\lambda_{min}$ of the equation $x^7+x^6+x^5+x^4-2x^2-x+1=0$ possessing a minimum absolute value. Solving this equation we get $\lambda_{min}=0.5698$. Consequently, the asymptotic behavior of interest can be expressed as $W_N^{im}= 1.134\cdot  1.7549^N$. This result is quoted in the main text.

We evaluate the number of all frozen states that should be extracted from the group {\bf II} in a similar manner.   All odd-numbered sequences within these states contain a single spin and all even-numbered sequences contain even numbers of spins or vice versa. Similarly to the previous consideration one can introduce numbers of spin states $Z_{N}^{oo}$, $Z_{N}^{eo}$, $Z_{N}^{oe}$, $Z_{N}^{ee}$ for states consisting of odd or even numbers of sequences (first superscript) and having one spin in odd- or even-numbered sequences (second superscript). One can introduce corresponding generating functions satisfying the equations below
\begin{eqnarray}
z_{oo}=xz_{eo}+2x, ~ z_{eo}=\frac{x^2}{1-x^2}z_{oo}, 
\nonumber\\
z_{oe}=\frac{x^2}{1-x^2}z_{ee}+\frac{2x^2}{1-x^2}, ~ z_{ee}=xz_{oe}. 
\label{eq:ImGrII}
\end{eqnarray}
 
 The generating function for the numbers of states of interest can be expressed as (cf. Eq. (\ref{eq:GrIIAns}))
 \begin{eqnarray}
 w_{II,im}=z_{eo}+z_{ee}+z_{oo}\frac{x^2+x^4}{(1-x^2)^2}=\frac{2x^3(2+x)}{1-2x^2}. 
 \label{eq:GrIIIm}
 \end{eqnarray}
 Consequently, the numbers of states  is defined by  the series expansion of Eq. (\ref{eq:GrIIIm}) as 
 \begin{eqnarray}
 W_N^{II, im} = 2^{N/2}\left[\frac{1+2\sqrt{2}}{4} +(-1)^{N}\frac{1-2\sqrt{2}}{4}\right] -\delta_{2,N}.  
 \label{eq:Gr2ImFin}
 \end{eqnarray}

\subsection{Counting group {\bf III} states} 
\label{sec:GenerFuncGrIII}

Here we evaluate the total number of spin  states having no frozen sets with the  structure $\{odd, 1, odd\}$, $\{odd,1,(even,1)_{k},odd\}$ ($k=1, 2,...$, see the main text).  We consider the problem for open boundary conditions (OBC). Problem with periodic boundary conditions (PBC) can be also solved using generating function method, but it is overcomplicated so we leave it to readers. The result for OBC gives the right asymptotic exponential dependence on the system size  for the number of states in the case of PBC and we estimate the preexponential factor by comparing the analytical asymptotic behavior with the numerical results. 

To evaluate the desirable number of states we use iteration procedure similar to that in the previous sections for the  numbers $W_{N}^{e}$ and $W_{N}^{o}$ of  spin states with the last sequence containing odd or even numbers of spins, respectively, with all mobile spins (except for the very last one in the last odd sequence)  and numbers $Y_{N}^{e}$, $Y_{N}^{o}$ for states ended by  even or odd spin sequences, respectively, with two or more spins in the end being frozen. Those states must have at the end the sequences $odd,(1, even)_{k}$ ($k=1,2,...$) for $Y_{N}^{e}$ or the sequence $odd,(1, even)_{k},1$ ($k=0,1,2,...$)  for $Y_{N}^{o}$ so the sequence in the end must contain one spin only. The first and last sequences in the state containing all mobile spins with OBC must have even numbers of spins  because the edge spin in the last odd sequence is always frozen. 

Then the iteration equations connecting the numbers of states with $N$ spins with the numbers of states with  smaller numbers of spins can be written as (cf. Eq. (\ref{eq:eveoe}))
\begin{eqnarray}
W_{N}^{e}=\sum_{k=1}^{\infty}(W_{N-2k}^{e}+W_{N-2k}^{o}+Y_{N-2k}^{e}), ~
W_{N}^{o}=\sum_{k=0}^{\infty}W_{N-2k-1}^{e}+\sum_{k=1}^{\infty}(W_{N-2k-1}^{o}+Y_{N-2k-1}^{e}),
\nonumber\\
 Y_{N}^{e}=\sum_{k=1}^{\infty}Y_{N-2k}^{o}+2\sum_{k=1}^{\infty}\delta_{N,2k}, ~
Y_{N}^{o}=Y_{N-1}^{e}+W_{N-1}^{o}.
\label{eq:shatter1}
\end{eqnarray}

Our target is the main exponential asymptotic for the total number of mobile spins $W_{m} \propto c\cdot A^{N}$. We wish to determine the factor $A$ analytically, while the prefactor will be estimated numerically. 

Similarly to the previous section, Eq. (\ref{eq:GenFunc}),  we introduce generating functions using low case letter names with the same subscripts as superscripts in capital letter notations. Eq. (\ref{eq:shatter1}) can be rewritten in terms of  generating functions as 
\begin{eqnarray}
\left(1-\frac{x^2}{1-x^2}\right)w_{e}-\frac{x^2}{1-x^2}w_{o}-\frac{x^2}{1-x^2}y_{e}=0, ~
-w_{e}\frac{x}{1-x^2}+w_{o}\left(1-\frac{x^3}{1-x^2}\right)-\frac{x^3}{1-x^2}y_{e}=0, 
\nonumber\\
y_{e}-\frac{x^2}{1-x^2}y_{o}=\frac{2x^2}{1-x^2}, ~
-xw_{o}-xy_{e}+y_{o}=0.
\label{eq:shatter1b}
\end{eqnarray}
Eq. (\ref{eq:shatter1b}) has the standard form of the linear equation $\widehat{M}\mathbf{x}=\mathbf{v}$ with the matrix $\widehat{M}$ defined as 
\begin{eqnarray}
\widehat{M}=
\begin{pmatrix}
1-\frac{x^2}{1-x^2} & -\frac{x^2}{1-x^2} & -\frac{x^2}{1-x^2} & 0 \\
-\frac{x}{1-x^2} & 1-\frac{x^3}{1-x^2}  & -\frac{x^3}{1-x^2}   & 0 \\
0 & 0 & 1 &  -\frac{x^2}{1-x^2} \\
0 & -x & -x & 1
\end{pmatrix}
\label{eq:MatrShat}
\end{eqnarray}
The solutions of Eq. (\ref{eq:shatter1b}) are inversely proportional to the determinant of this matrix. They have poles $1/(x-\lambda)$ at zeros $\lambda$ of that  determinant. The large $N$ asymptotic of the solution $W_{No,e}$ can be expressed using the  root $\lambda_{min}$ having  the smallest absolute value in the form   $W_{No,e} \propto \lambda^{-N}$. This can be proved using the partial fraction expansion similarly to previous sections. 

The determinant of the matrix $\widehat{M}$ can be expressed as 
\begin{eqnarray}
{\rm det} \widehat{M}=\frac{4x^5-2x^4-3x^3-3x^2+1}{(1-x^2)^2}. 
\label{eq:DetShat}
\end{eqnarray} 
We found the solutions of the equation det $\widehat{M}=0$ numerically. The solution with the minimum absolute value is $\lambda_{min} \approx 0.5423$. Consequently, one can expect the number of states without frozen spins to scale as $W_{N} \propto \lambda_{min}^{-N} = 1.8442^{N}$. This number increases faster then the numbers of states in groups {\bf I} and {\bf II} and, therefore, subtracting their numbers will not affect the main exponential asymptotic.  

The solution for the generating function of interest $w_{e}(x)$ representing the states with all mobile spins takes the form 
\begin{eqnarray}
w_{e}(x)=\frac{x^4}{(4x^5+2x^4-3x^3-3x^2+1)}. 
\label{eq:MobDer}
\end{eqnarray}
Expanding this result into simple fractions one can define the preexponential factor in the dominating exponential asymptotic  for the OBC problem that is $W^{e}_{N}=0.0551\cdot 1.8442^{N}$

To derive the exponential asymptotic for the number of group {\bf III} states in the periodic system we compare analytical results vs. numerical calculations of the numbers of states having no frozen spins as reported in Fig. \ref{fig:GrIII}. It turns out that the best exponential fit can be obtained for the total number of states belonging to groups {\bf I}, {\bf II} and {\bf III} and this fit is $W^{e}_{N}=1.0 \cdot 1.8442^{N}$. The number of group {\bf III} states approaches this asymptotic with increasing the number of spins as it supposes to be because the relative weight of states of groups {\bf I} and {\bf II} decreases exponentially with increasing the number of spins compared to that of the group {\bf III}. 

\begin{figure}
\includegraphics[scale=0.4]{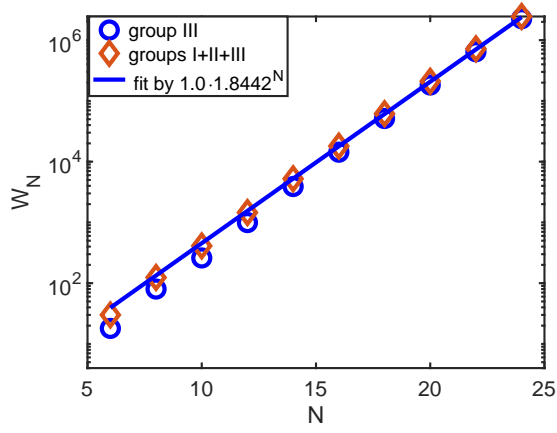} 
\caption{Numbers of states with all mobile  spins vs the number of spins  compared to the analytical  fit  $1.0\cdot 1.8442^{N}$}
\label{fig:GrIII}
\end{figure}

\section{Imbalance as a measure of system dynamic properties} 
\label{sec:Imbal}

Here we consider the definition of an infinite time imbalance in greater detail with special attention to its experimental determination. 
The imbalance is defined following the conditions of a typical experiment similarly to  Ref. \cite{Monroe2021observationStarkMBL} (Eq. (\ref{eq:ImbDefMT}) in the main text).  It is determined by  the system evolution from some initial product state $|a>$, characterized by fixed projections of spins $\pm 1/2$ to the $z$ axis ($t=0$). The system evolution  can be expressed in terms of average spin projections $<S_{k}^{z}(t)>$ at time $t$. 
The extraction of the infinite time limit for $<S_{k}^{z}(t)>$   is not straightforward  using  the experimental data available at a finite time only. Yet it can be estimated reasonably well as described below. 



According to our observations imbalance converges to its infinite time limit quite fast   for the initial state belonging to the groups {\bf I} or {\bf II} with delocalized eigenstates. For the family of initial states belonging to the group {\bf II}, shown in Fig. \ref{fig:ImbTDepPr}.{\bf a},  imbalances approach its  infinite time limit during the time of order of few inverse spin hopping amplitudes $\Delta^{-1}$ Eq. (\ref{eq:MinimHamCons}). The convergence gets better  with increasing the number of spins as illustrated in Fig.  \ref{fig:ImbTDepPr}.{\bf a},. 

For the localized state considered in Fig. \ref{fig:ImbTDepPr}.{\bf b}, originated from the group {\bf III} state,  shown in the graph,   the oscillations of imbalance around its infinite time limit are very strong. However, one can estimate   the  infinite time limit of imbalance using time averaged imbalance defined as  
\begin{eqnarray}
I_{av}(t)=\frac{\int_{0}^{t}I(\tau)d\tau}{t}.
\label{eq:ImbtavDef}
\end{eqnarray}
According to Fig.  \ref{fig:ImbTDepPr}.{\bf b} the time averaged imbalance  $I_{av}(t)$ converges to its infinite time limit after the time around few inverse hopping amplitudes $\Delta^{-1}$ similarly to Fig.   \ref{fig:ImbTDepPr}.{\bf a}. Thus  the infinite time limit of imbalance can be determined  experimentally and compared with the predictions of theory.  One should notice that this definition can be not applicable  in marginal situations of very slow dynamics emerging for instance for unlikely realized states of anomalously high or low energy \cite{ab2017comment}.

\begin{figure}[h]
\centering
\subfloat[]{\includegraphics[scale=0.4]{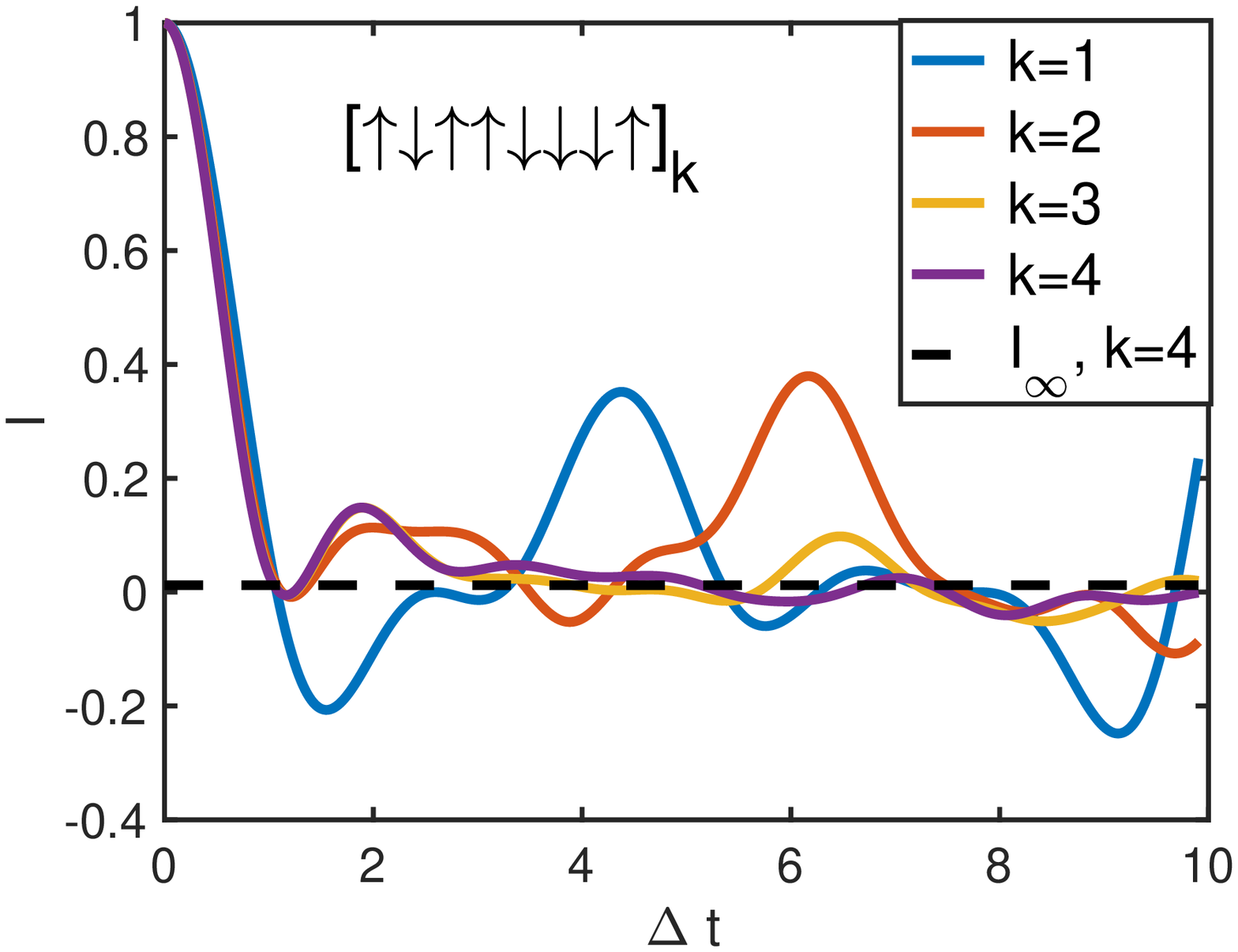}}
\subfloat[]{\includegraphics[scale=0.4]{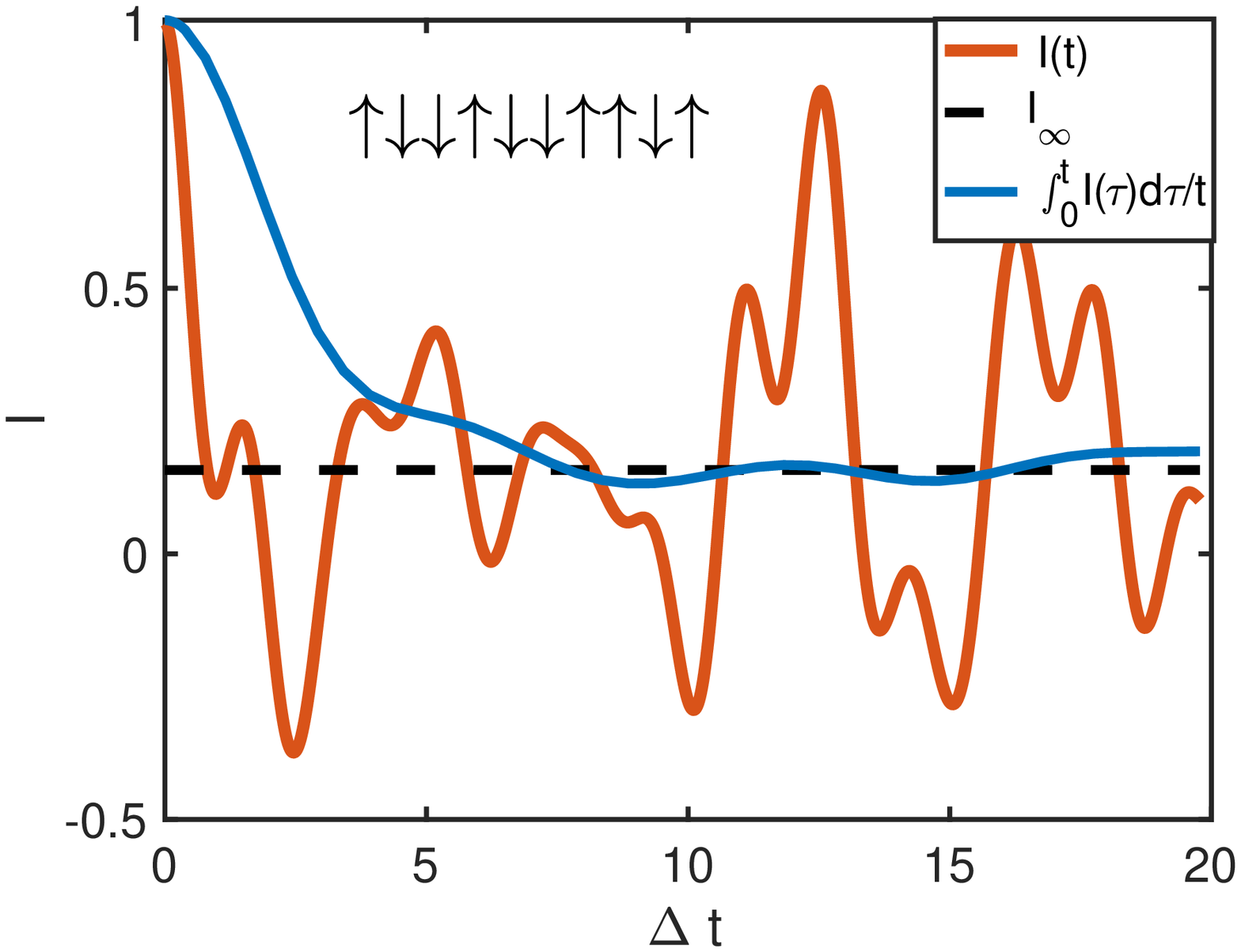}}
\caption{\small   ({\bf a}) Convergence of the imbalances for the family of states belonging to the group {\bf II} to the infinite time limit. ({\bf b}) Convergence of the time averaged imbalance to its infinite time limit for the localized state belonging to the group {\bf III}. The initial product states are shown within the graphs in the inverted representation. 
} 
\label{fig:ImbTDepPr}
\end{figure}

Eq. (\ref{eq:ImbDefMT}) determines the imbalance measured in Ref. \cite{Monroe2021observationStarkMBL} and we use it here for a characterization of different groups of states. The Hamming distances reported  in Ref. \cite{Guo2020MBLTiltedModel} can be expressed as 
\begin{eqnarray}
HD(t)=1-\frac{4}{N}\sum_{k=1}^{N}<S_{k}^{z}(t)><S_{k}^{z}(0)>.
\label{eq:HamDistDefMT}
\end{eqnarray}
In a thermodynamic limit of $N\rightarrow \infty$ one can set $<S^{z}>=0$ in the definition of imbalance (Eq. (\ref{eq:ImbDefMT}) in the main text). Then two definitions becomes equivalent, since $HD(t)+I(t)=1$. However, we prefer to use Eq. (\ref{eq:ImbDefMT}) or its generalizations above since they are  less sensitive to finite size effects due to the subtraction of the finite size expectation values for average spin projections. 

\end{widetext}

\end{document}